\documentclass[noeprint,twocolumn,showpacs,amsmath,amssymb,sort&compress,floatfix,superscriptaddress,prl,aps,noeprint]{revtex4-2}

\usepackage{graphicx}
\usepackage{dcolumn}
\usepackage{bm}
\usepackage[hidelinks]{hyperref}
\usepackage[mathlines]{lineno}
\usepackage{mathtools}
\usepackage{amsmath}
\usepackage{float}
\usepackage[caption=false]{subfig}
\usepackage[english]{babel}
\usepackage[usenames,dvipsnames]{xcolor}
\usepackage{xparse}
\usepackage{lipsum}
\usepackage{afterpage}
\usepackage{capt-of}
\newcommand*{\balancecolsandclearpage}{%
  \close@column@grid
  \clearpage
  \twocolumngrid
}

\DeclareMathOperator{\sinc}{sinc}
\DeclareMathOperator{\Li}{Li}
\let\Re\relax
\let\Im\relax
\DeclareMathOperator{\Re}{Re}
\DeclareMathOperator{\Im}{Im}

\usepackage{etoolbox}

\makeatletter
\renewcommand{\fnum@figure}{FIG. \thefigure}
\makeatother

\newcommand{\bl}[1]{{\color{black}#1}}
\newcommand{\tns}[1]{{\color{black}#1}}

\newcommand{\ue}{SUPA, School of Physics and Astronomy, University of Edinburgh, Peter Guthrie Tait Road, Edinburgh EH9 3FD, United Kingdom}

\begin{document}

\title{Passive Janus Particles Are Self-propelled in Active Nematics}

\author{Benjamin Loewe,}
\email[Corresponding author \\]{bloewe@ed.ac.uk}
\affiliation{\ue}
\author{Tyler N. Shendruk}
\affiliation{\ue}


\begin{abstract}
While active systems possess notable potential to form the foundation of new classes of autonomous materials \cite{Zhang2021}, designing systems that can extract functional work from active surroundings has proven challenging. In this work, we extend these efforts to the realm of designed active liquid crystal\tns{/colloidal} composites. We propose suspending colloidal particles with Janus anchoring conditions in an active nematic medium. These passive Janus particles become effectively self-propelled once immersed into an active nematic bath. The self-propulsion of passive Janus particles arises from the effective $+1/2$ topological charge their surface enforces on the surrounding active fluid. We analytically study their dynamics and the orientational dependence on the position of a companion $-1/2$ defect. 
We predict that at sufficiently small activity, the colloid and companion defect remain bound to each other, with the defect strongly orienting the colloid to propel either parallel or perpendicular to the nematic. At sufficiently high activity, we predict an unbinding of the colloid/defect pair. This work demonstrates how suspending engineered colloids in active liquid crystals may present a path to extracting activity to drive functionality. 
\end{abstract}

\maketitle

\section{Introduction}

Active matter extracts energy from internal or external sources and transforms it into persistent motion \cite{Marchetti2013}. Although many physical realizations occur in living systems, such as bacterial colonies \cite{Dombrowski2004, Zhang2010, Wioland2016}, cellular tissues \cite{Serra-Picamal2012, Blanch-Mercader2017}, flocking animals and human crowds \cite{Bain2019}, there has also been considerable success developing and understanding synthetic active systems. Examples include swarming bots \cite{Giomi2013}, vibrated granular matter \cite{Deseigne2010}, self-propelled nanorods \cite{Paxton2004,Liu2011}  and active colloidal particles \cite{Valadares2010, Bricard2013, Palacci2013}. In each of these examples, the work needed to achieve motility comes from either chemical reactions or responses to external fields.

\tns{An alternative route is to immerse a passive object into an out-of-equilibrium system, extract work out of the environment \cite{Sokolov2010, Ekeh2020, Holubec2020}, and transform it into a driving force \cite{Angelani2010, Kaiser2014, Yan2018, Pietzonka2019, Ramos2020, Belan2021, Decayeux2021}. In previous work, \bl{passive} particles were suspended into motile bacterial baths \cite{wu2000, Peng2016, Patteson2016, Ortlieb2019}. However, only rarely has it been explored how active suspensions can power motility of passive objects \cite{Sanchez2012, Mallory2014, Laskar2015, Thampi2016, Zhang2021}. Active nematics \cite{Gruler1999, Ramaswamy2003, Ahmadi2006, Narayan2007,Sanchez2012,Keber2014, Kawaguchi2017, Saw2017, Ellis2018}} possess many appealing properties for this task. Active nematic films are characterized by the presence of point disclinations in which the orientational order vanishes and around which the nematic director rotates \tns{by} an angle $\pm \pi$. Because this winding cannot be untangled by smooth variations of the director, these defects are \tns{topological in nature, carrying} a topological charge of $\pm 1/2$ (higher charge defects are energetically unfavorable). However, in contrast with passive nematics, a $+1/2$ defect in an active nematic system generates spontaneous flows that yield non-zero advection at the defect core. This makes +1/2 defects motile \cite{Narayan2007, Giomi2013a}. The self-propulsion of these defects is sufficient to overcome the attractive interaction \bl{exerted} by neighbouring non-motile $-1/2$ defects, allowing unbinding of defect pairs\tns{,} which drives the system into a chaotic state known as active turbulence \cite{Shankar2018}.  

\begin{figure}
    \centering
    \includegraphics[width=\columnwidth]{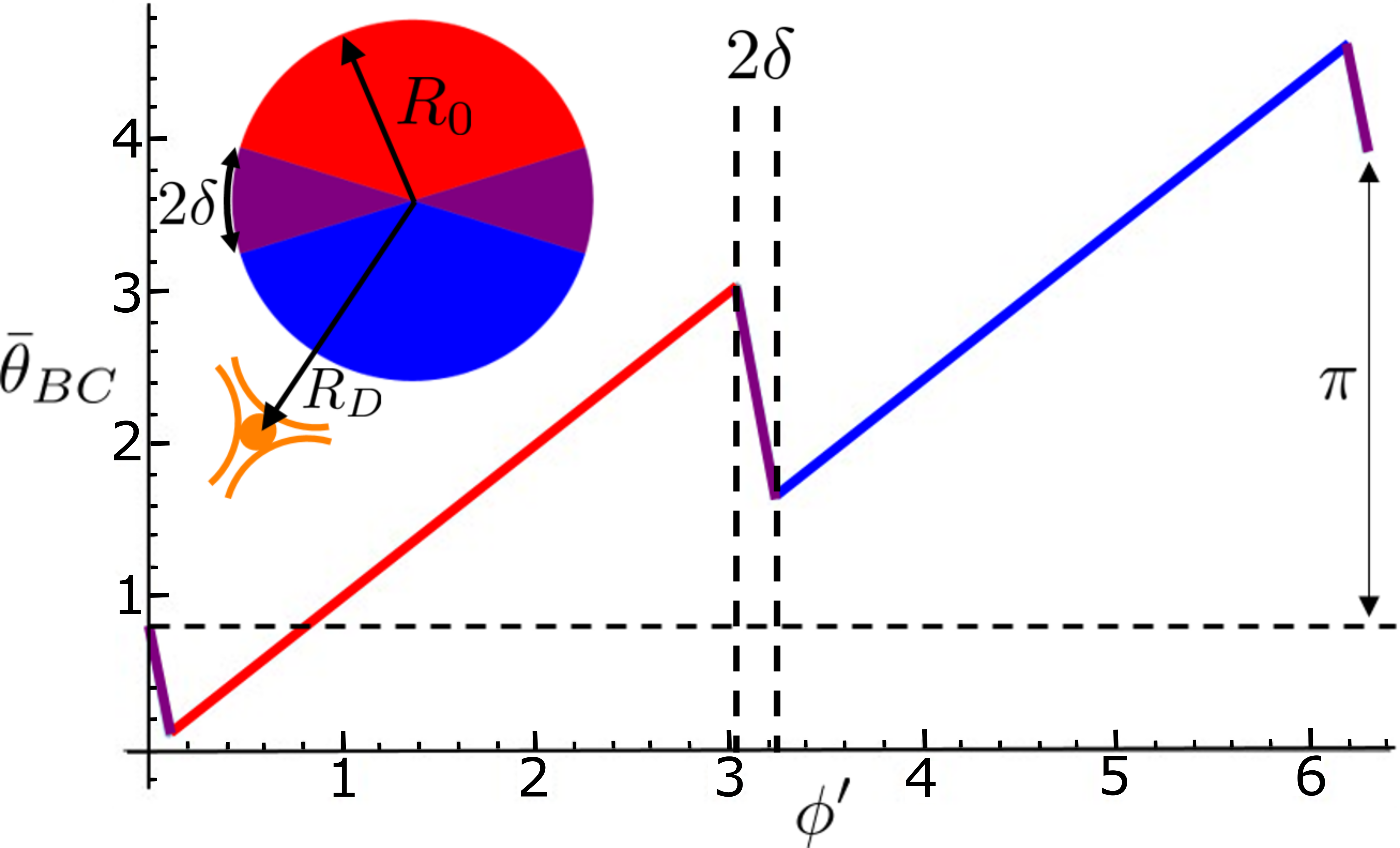}
    \captionof{figure}{The $\bl{\bar{\theta}_{BC}}$ contribution to the anchoring condition for a Janus colloid of radius $R_0$, see Eq.~(\ref{eq:BC2})  The red \tns{semicircle} represents homeotropic anchoring, while the blue pole is planar with continuous transition zones of width $2\delta$ at the equator that soften the $\pi/2$ jump between homeotropic and planar anchoring (purple wedges). Notice the finite jump of $\pi$ between $\phi'=0$ and $\phi'=2\pi$, which provides the colloid with an effective topological charge of $+1/2$.  On the insert, we can find, with matching colors, the different regions of anchoring over the colloid. At a distance $R_D$ sits the companion $-1/2$ defect, here depicted in orange.}
    \label{fig:BC}
\end{figure}

Another important property of nematics is that they can anchor to surfaces with a prescribed orientation \cite{Stark2001}. If the anchoring is strong enough, this prescribed orientation leads to a non-zero winding of the nematic director and thus to an effective topological charge localized at the interior of the colloid \cite{Terentjev1995, Ramaswamy1996, Stark2001, Dogic2014}. \tns{Colloids with strong, uniform homeotropic or planar anchoring have an effective charge of $+1$, which must be balanced by associated defects in the nematic surroundings or at the colloid surface}.

\begin{figure*}
    \centering
    \includegraphics[width=\textwidth]{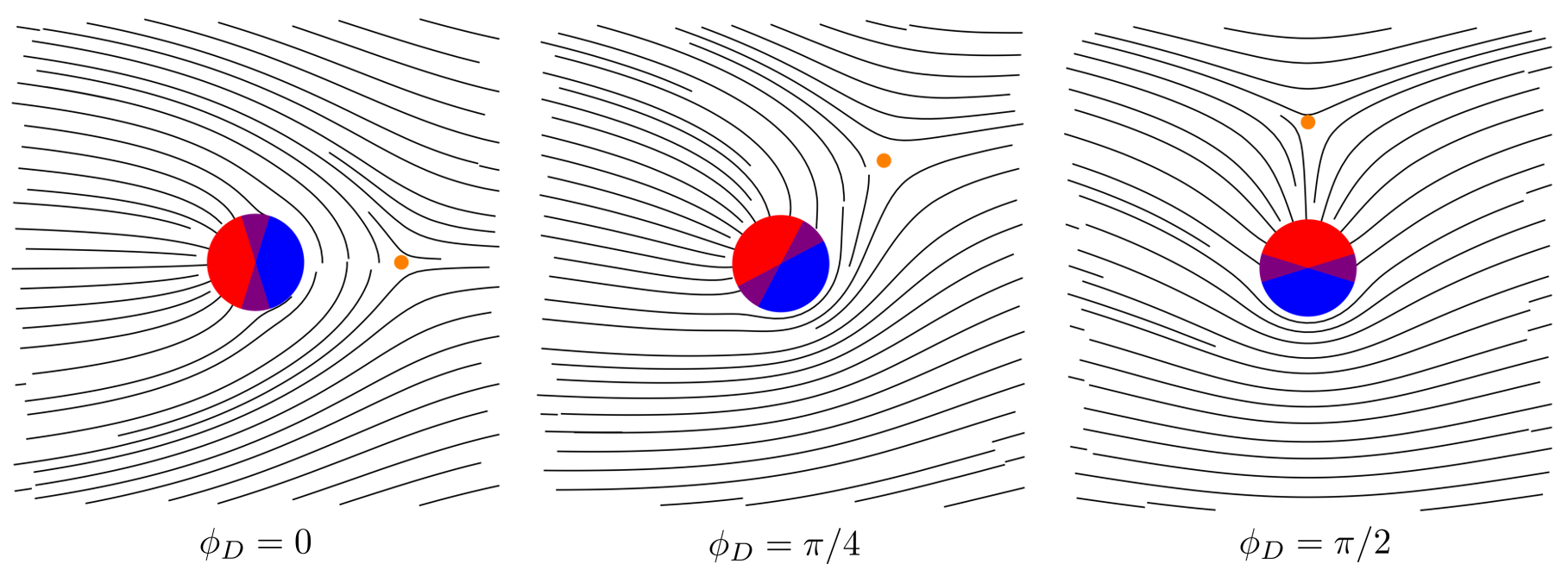}
    \captionof{figure}{Nematic field around the Janus colloid according to Eqs.~(\ref{eq:solution1} and \ref{eq:solution2}) for different angular positions \tns{$\phi_D$} of the companion $-1/2$ defect. The sum was cut-off at $N=100$ and $R_D=3R_0$. The orange dots mark the position of the $-1/2$ defect. \tns{The solution is for the combination of defect location $\phi_D$ and colloid orientation $\phi_c$ that together minimize the free energy. Therefore, setting the defect position $\phi_D$ determines the colloid orientation $\phi_c$. Varying the set defect angle results in a counter-rotation of the colloid orientation.}}
    \label{fig:solutions}
\end{figure*}

Given both the motility  of $+1/2$ disclinations in active nematics and the possibility of creating effective topological defects through the inclusion of colloids, it is natural to ask if it is possible to design a two-dimensional colloid with \tns{topological charge} $+1/2$ and if such colloid acts \tns{as an effectively self-propelled particle}. We focus in 2D because most active nematics occur in 2D films \cite{Sanchez2012,Ellis2018,Rivas2020}. In this paper, we demonstrate that such self-propulsion is possible by designing a colloid with a Janus structure. Our Janus colloid possesses homeotropic anchoring on one \tns{semicircle (red pole in figures)}, planar anchoring on the other \tns{(blue pole)}, and continuous transition zones at the equator, resulting in an effective $+1/2$ topological charge in the system. \bl{In analogy with positive $+1/2$ topological defects, the Janus colloid is self-propelled. The resulting colloidal self-propulsion is similar to that of active Janus particles \cite{Ebbens2010} or Quincke rollers \cite{Bricard2015}, in that our passive colloid extracts energy from its surroundings to achieve propulsive motion}. Conservation of topological charge requires that each Janus colloid is accompanied by a $-1/2$ topological defect in the surrounding nematic. 

We employ analytical \tns{models} to construct such a colloid, estimate its self-propulsion and characterize its dynamics. By estimating and integrating the total stress at the colloid's surface, we learn that the \tns{surrounding activity does indeed drive} a net non-zero propulsive force. Thus, the passive Janus particle effectively behaves as a \bl{self-propelled particle}.  Crucially, we show that the self-propulsion of the colloid, at the low activity limit, is primarily parallel or perpendicular to the direction of local nematic order. On the other hand, high activity leads to spontaneous decoupling of the colloid with its companion \tns{defect}. 

This novel activity-driven method of self-propulsion sheds light in how to extract work from active nematics and may serve as a new direction for collective phenomena. As these colloids are topologically charged, they experience long-range elasticity-mediated dipolar interactions which could lead to flocking or other collective behaviours. Furthermore, combining these systems with light activated molecular motors \cite{Zhang2019} could open the door to activity gradients and a possible control mechanism for colloid self-propulsion through active materials.



\section{Model}

\begin{figure*}
    \centering
    \includegraphics[width=\textwidth]{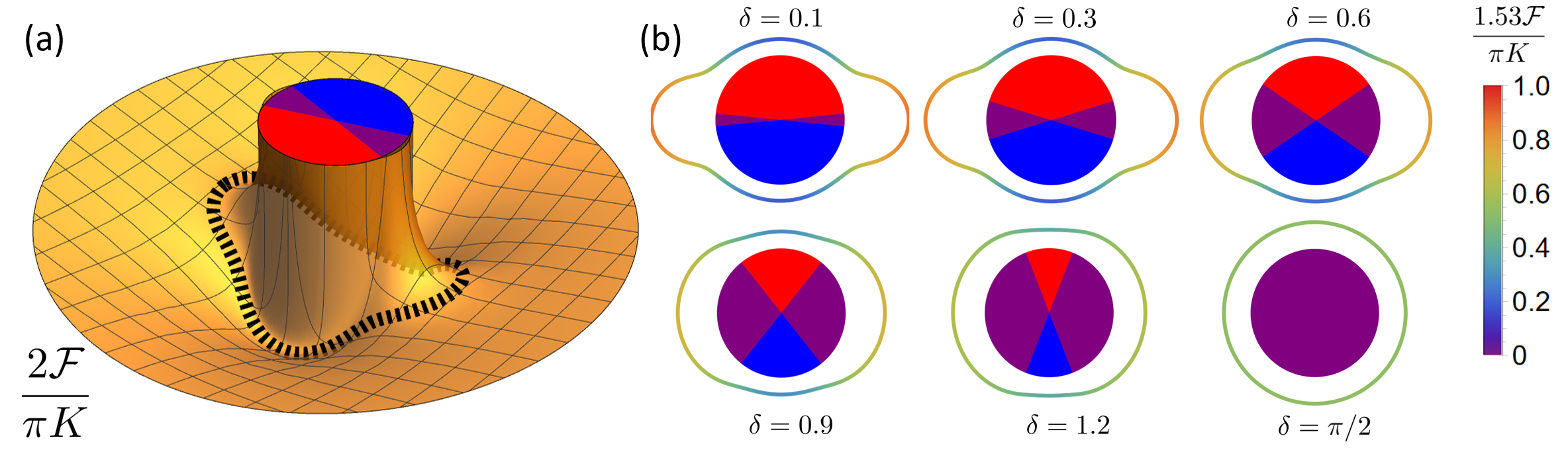}
    \captionof{figure}{(a)  Free energy landscape as surrounding a Janus colloid ($\delta = 0.3$) as a function of the position of the companion defect. The dotted line denotes the curve of stationary radii $r_s$(\emph{i.e.}, points at which $\partial_r\mathcal{F} =0$). Its shape is a result of the competition between the long-range \bl{attractive} force between opposite topological charges, the short-range repulsion due to the strong anchoring condition and the repulsion between the negative topological charge density distributed along the transition zones.  Notice how the landscape has its minima on the poles, which are the furthest away from the transition zones. In contrast, on the equator the landscape only has a local maximum. In agreement with the above, we also see the stationary radii in the equator are further away of the colloid than in the poles, leading to a lemon-shaped curve. (b) Curve of stationary radii for different values of $\delta$; the coloring of the curves correspond to the local value of $\mathcal{F}$. In agreement with our interpretation of a locally distributed negative charge density on the transition zones, as these become wider and occupy a larger sector of the colloid, their repulsion becomes less pronounces leading to a more isotropic curve, and finally a circle for $\delta=\pi/2$.}
    \label{fig:stat}
\end{figure*}
We consider a single colloid of radius $R_0$ centered at the origin, immersed in a two-dimensional nematic that is globally aligned along the $x$-axis far from the colloid. \tns{For the moment, the nematic can be treated as passive, but we will consider non-zero activity further below}. For simplicity, we work in the far-field approximation.  This implies our model is a long length-scale theory in which the size of a defect core acts as a natural cut-off. As such, the alignment of the nematic liquid crystal is entirely described through its director $\hat{\boldsymbol{n}} = \cos(\theta)\hat{\boldsymbol{e}}_x+\sin(\theta)\hat{\boldsymbol{e}}_y$ via the nematic orientation $\theta$. Furthermore, we will work in the one-Frank-constant approximation, and thus the energetic cost of bending or splaying the director is given by

\begin{equation}
\label{free_energy}
\mathcal{F} = \frac{K}{2}\int d^2\bm{r} |\boldsymbol{\nabla}\theta|^2, 
\end{equation}
where $K$ denotes the Frank constant.

The colloid imposes an anchoring condition for the nematic at its surface, denoted by $\theta_{BC}(\phi)$. This function is: i) not constant, and ii) not continuous,  since the nematic director is defined modulo $\pi$. We discuss the physical consequences of such discontinuities further below. Moreover, we work in the strong anchoring limit, in which it is energetically prohibitive for the director to deviate from the prescribed anchoring. Under these conditions, the anchoring can be treated as a boundary condition for the nematic orientation, $\theta(r=R_0, \phi) = \theta_{BC}(\phi)$.  This results in the nematic director winding around the colloid surface \tns{$k = (\theta_{BC}(2\pi)-\theta_{BC}(0))/(2\pi)$ times}, producing an effective topological charge $k$.

Because the system is globally aligned, the total topological charge of the system must remain zero; therefore, the colloid must induce \tns{one or more} companion topological defects whose topological charge must sum to $-k$. In polar coordinates centered in the colloid, the position of the $i$-th defect is ($R^i_D$,$\phi^i_D$). Their angular position $\phi_D$ matches the angular position of the discontinuities in the anchoring boundary condition $\theta_{BC}$, with the defects' topological charge being proportional to the size of the jump in the discontinuity. \tns{Here, we consider a colloidal design which necessitates only a single $-1/2$ defect.} 

 Moreover, we work in an adiabatic approximation in which the relaxation time for the nematic director is assumed microscopic. As such, the system is invariably at its energy minimum. Minimizing the free energy, $\delta\mathcal{F}/\delta\theta=0$, we obtain
\begin{equation}
\label{eq:lapplace}
    \nabla^2\theta = 0,
\end{equation}
contingent on $\theta(r=R_0,\phi) =\theta_{BC}(\phi;\phi_D,\phi_c)$ and $\theta(r\rightarrow\infty,\phi) = 0$, where $\phi_c$ is the orientation of the colloid. Our description does not explicitly account for contributions due to the orientation of the companion defect \cite{Vromans2016, Tang2019} --- the defect orientation is that which corresponds to the minimized free energy. \tns{Likewise, if the angular position $\phi_D$ of the defect is set then the colloidal orientation $\phi_c$ must minimize the free energy.} Further details on the boundary conditions of $\theta$ are provided in \tns{A}ppendix A.

\section{Janus Boundary Condition}

Inspired by the motility of $+1/2$ defects \cite{Narayan2007, Giomi2013a}, we choose a boundary condition with an effective topological charge of $+1/2$.  We divide the colloid \tns{into two semicircles}, one with homeotropic anchoring \tns{(red)} and the other with planar anchoring \tns{(blue)}, with two continuous transition zones, each of angular width $2\delta$, at the equator (see Fig.~\ref{fig:BC}). The role of such transitions zones is to smooth the $\pi/2$ jump between homeotropic and planar anchoring, which cannot happen discontinuously since such a jump would violate nematic symmetry.  In the reference frame of the colloid, defined by $\phi_c=0$, this boundary condition takes the form
\begin{equation}
\label{eq:BC}
    \theta_{BC}(\phi') = \bar{\theta}_{BC}(\phi')-\pi\, \Theta(\phi'-\phi'_D),
\end{equation}
where $'$ denotes angles in the colloid reference frame with respect to the colloidal equator. 
\tns{The connection between the colloid reference frame and the lab frame is illustrated in Appendix A, Fig.~\ref{fig:rotation}.} We define
\begin{equation}
\label{eq:BC2}
    \bar{\theta}_{BC}(\phi')\equiv 
    \begin{cases}
        \frac{\pi}{4}+\left(1-\frac{\pi}{4\delta}\right) \phi' & \phi'<\delta \\
        \phi' & \delta \leq \phi' < \pi-\delta \\
        \frac{\pi(\pi-\delta)}{4 d}+\left(1-\frac{\pi}{4\delta}\right)\phi' & \pi-\delta \leq \phi' < \pi+\delta \\
        \phi'-\frac{\pi}{2} & \pi+\delta \leq \phi' < 2\pi-\delta\\
        \frac{\pi(2\pi-3\delta)}{4\delta}+\left(1-\frac{\pi}{4\delta}\right)\phi' & 2\pi-\delta < \phi'
    \end{cases},
\end{equation}
and $\Theta(x)$ denotes the Heaviside function.  The $\bar{\theta}_\text{BC}$ term sets the director across all regions of the colloid (Fig.~\ref{fig:BC}), while the Heaviside function in Eq.~(\ref{eq:BC}) introduces a $-\pi$ jump at $\phi'_D$, which is physically allowed under head-tail symmetry.  Mathematically\tns{,} however, this introduces a branch cut which identifies the position of the companion defect. In the limit $\delta\rightarrow \pi/2$ (\emph{i.e.}, when the entire colloid is covered by transition zones), $\bar{\theta}_{BC}(\phi') = \pi/4+\phi/2$, identical to the form that $\theta$ acquires around an isolated $+1/2$ defect.  As such, our choice for the boundary condition also allows us to explore the properties of a colloid that perfectly imitates a $+1/2$ defect.

We solve \bl{Eq.}~(\ref{eq:lapplace}) for this \bl{choice} of $\theta_{BC}$ in Appendix A. \bl{The solutions are obtained by writing general forms of the interior and exterior solutions to the Laplace equation and imposing appropriate boundary conditions to obtain explicit forms for the coefficients of the solution.} Crucially, since the total topological charge enclosed by concentric radius changes from $+1/2$ to $0$ \tns{at the location of} the companion defect, we must separate our domain \tns{into two} regions: Region I, from the surface to the companion defect ($R_0\leq r<R_D$), and region II, beyond the companion defect ($r>R_D$) \bl{see Fig.~\ref{fig:BC2}(d)}. Our solution reveals that the polarization of the colloid, $\phi_c$, and the position of the companion defect are linked in a one-to-one relationship, $\phi_c = \pi/4-\phi'_D/2$ \bl{(see Eq.~\ref{eq:theta_II} and Fig.~\ref{fig:rotation} in Appendix A)}. 
This one-to-one relationship implies that \tns{re-orienting} the companion \tns{defect's position} induces \tns{a change in the colloid orientation to minimizes the free energy} \cite{Vromans2016, Tang2019}. As we have not set the colloid or the companion defect orientation \emph{a priori}, our solution selects the combination that minimizes the free energy for \tns{each} given configuration. The colloid's orientation dependence on the lab-frame defect angular position, $\phi_D$, is given by $\phi_c = (\pi-2\phi_D)/2$, which tells us that as the \tns{set defect position is moved, the equilibrium colloid orientation will counter-rotate}.  Although this effect is apparently independent of the separation, \tns{the} colloid must move a distance comparable to $R_D$ in order for $\phi_D$ to change considerably. This shows that the effect actually weakens with distance. Having establish this connection, we can write the solution to Eq.~(\ref{eq:lapplace}) in the lab frame\bl{ (see Appendix A)} \tns{to be}
\begin{widetext}
\begin{align}
\label{eq:solution1}
    \begin{split}
        \theta_I(r,\phi) = {}&  \frac{1}{2}\left(\phi+\phi_c+\frac{\pi}{2}\right)-\pi \Theta(\phi-\phi_D)\\
        &-\sum_{n=1}^{\infty}\left[\frac{(1+(-1)^n) \sin(n \delta)\sin(n(\phi-\phi_c)) }{2 \delta n^2}\left(\frac{R_0}{r}\right)^n\right.
        \left.+\frac{\sin(n(\phi-\phi_D))}{2 n}\bl{\left\{\left(\frac{R_0^2}{R_Dr}\right)^n-\left(\frac{r}{R_D}\right)^n\right\}}\right],
    \end{split}\\
    \label{eq:solution2}
    \begin{split}
        \bl{\theta_{II}(r,\phi)}  = {}&
       -\sum_{n=1}^{\infty} \tns{\left[\frac{(1+(-1)^n) \sin(n \delta)\sin(n(\phi-\phi_c)) }{2 \delta n^2}\left(\frac{R_0}{r}\right)^n \right.
        \left.+\frac{\sin(n(\phi-\phi_D))}{2 n}\left\{\left(\frac{R_0^2}{R_Dr}\right)^n + \left(\frac{R_D}{r}\right)^n\right\}\right]},
    \end{split}
\end{align}
\end{widetext}
in which $\theta_I$ and \bl{$\theta_{II}$} denote the solution in regions I ($R_0\leq r<R_D$) and II ($r>R_D$), respectively. 

The first two terms in Eq.~(\ref{eq:solution1}) for $\theta_I(r,\phi)$ provide both the orientation of the colloid and the behaviour of the nematic in the intermediate regime $R_0 \ll r \ll R_D$\tns{. In this region,} all other terms are negligible. Not surprisingly, this corresponds to a nematic surrounding an isolated $+1/2$ defect.  The first term inside the sum in Eq.~(\ref{eq:solution1}) describes the behaviour of the nematic in the close vicinity of the colloid and directly reflects the prescribed anchoring.  The second term inside the sum describes the deformation of the nematic field due to the presence of the companion defect. Regarding the form of $\theta_{II}$, its terms play the same role that their counterpart in $\theta_I$, with the distinction that the first term in the sum\bl{, which goes as $\sim (R_0/r)^n$ with $r>R_D$,} only becomes relevant if $R_D$ is comparable with $R_0$, \emph{i.e.}, when the companion defect resides in the vicinity of the colloid surface. 

Figure~\ref{fig:solutions} depicts the solution of  Eqs.~(\ref{eq:solution1}-\ref{eq:solution2}) for representative companion defect positions when $R_D=3R_0$. In Fig~\ref{fig:solutions}(a), $\phi_D=0$\tns{, such that} the colloid/defect orientation is aligned parallel to the global nematic orientation. In this configuration, the colloid is also oriented parallel to the far-field. Because the planar-anchored pole \tns{(blue)} is closest to the defect, the elastic deformation between the surface and the defect is primarily bend. As the defect position is moved to $\phi_D=\pi/4$, the \tns{equilibrium colloid orientation must turn} clockwise to compensate (Fig~\ref{fig:solutions}(b)). 
Because of this, the defect is closest to the equator in this angular position. This process continues as we move the defect position to $\phi_D=\pi/2$. By this point, the \tns{equilibrium colloid orientation} has rotated $-\pi/2$; thus, the colloid/defect complex is now perpendicular to the far-field director.
The companion defect is now closest to the homeotropic pole \tns{(red)} and so the principle deformation mode between the colloid and the surface is splay. 

\begin{figure}[tb]
    \includegraphics[width=\columnwidth]{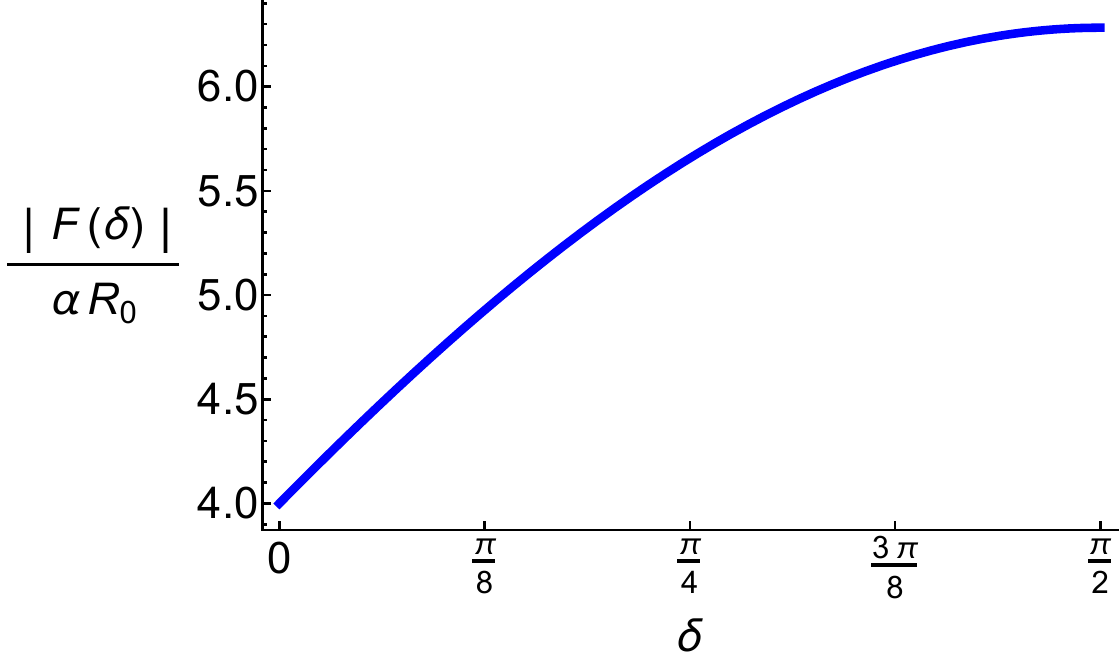}
    \caption{Magnitude of the propulsive force as a function of $\delta$. The magnitude is an increasing function of $\delta$, growing linearly at first, then saturating and reaching a plateau near $\delta=\pi/2$.}
    \label{fig:force_mag}
\end{figure}

\section{Colloid-Defect Interaction}

The interaction energy between the colloid and its companion defect can be obtained by inserting Eqs.~(\ref{eq:solution1}) and (\ref{eq:solution2}) in Eq.~(\ref{free_energy}) (see Appendix B). The result, in the colloid reference frame, is
\begin{equation}
    \label{eq:FE1}
    \begin{split}
        \frac{\mathcal{F}}{K} &= \frac{\pi}{2}\ln\left(\rho\right)- \frac{\pi}{4}\ln\left(1-\frac{1}{\rho^2}\right)\\
        &+\frac{\pi}{8\delta} \Im\left[\Li_2\left(\frac{e^{2 i(\phi'_D+\delta)}}{\rho^2}\right)-\Li_2\left(\frac{e^{2 i(\phi'_D-\delta)}}{\rho^2}\right)\right]
    \end{split}
\end{equation}
in which we have defined the dimensionless coordinate $\rho=R_D/R_0$ and $\Li_2$ denotes the dilogarithm.  The term $\sim \ln(\rho)$ corresponds to the attractive interaction between $+1/2$ and $-1/2$ defects.  Since this term is identical to a $+1/2$ defect in the far field, it can be thought of as the leading term in a multipole expansion of the interaction.  The second term, which \tns{goes as} $\sim \ln(1-\rho^{-2})$, is a short-range repulsion between the colloid and the defect, \tns{and} is logarithmically divergent. It arises because of our assumption of a strong anchoring: In contrast to an ``actual'' +1/2 singularity, the anchoring at the surface of the colloid does not change regardless of how close the defect is, leading to an increase of elastic energy. As such, at close distances the colloid behaves as a wall, strongly repelling the defect.  The third term corresponds to the remaining contributions of the multipole expansion and contain the details of the interaction at intermediate distances.

As it can be seen in Fig.~\ref{fig:stat}(a), $\mathcal{F}$ is highly anisotropic. The colloid repels the defect more strongly at the transition zones. This can be seen in Fig.~\ref{fig:stat}(b), which depicts the curves of radial stationary points at which $\partial_r\mathcal{F}=0$. The repulsion weakens and becomes more uniform as the transition zone covers a larger sector of the colloid's surface. If the transition zone covers the entire colloid, then the repulsion is completely uniform. This behaviour originates due to the presence of a negative topological charge density distributed along the surface of the transition zones (see Appendix B).  Indeed, Eq.~(\ref{eq:FE1}) can be \tns{recast as} 
\begin{equation}
\label{eq:FE2}
\begin{split}
    \frac{\mathcal{F}}{K} = & \frac{3\pi}{2}\ln\left(\rho\right)- \frac{\pi}{4}\ln\left(1-\frac{1}{\rho^2}\right)\\
    &-\frac{\pi}{4\delta}\left[\int_{-\delta}^{\delta}d\beta \ln\left(\tilde{\rho}(\beta,\phi'_D,\rho)\right)\right.\\
    &\qquad\left.+\int_{\pi-\delta}^{\pi+\delta}d\beta \ln\left(\tilde{\rho}(\beta,\phi'_D,\rho)\right)\right] ,
    \end{split}
\end{equation}
where 
\begin{equation}
\tilde{\rho}(\beta,\phi',\rho) = \sqrt{1+\rho^2-2\rho\cos(\phi'_D-\beta)}
\end{equation}
is the distance between the defect and an element of topological charge $-d\beta/(4\delta)$ at the surface of the transition zone in units of $R_0$. This negative charge density integrates to a total of -1, and is superposed with a positive charge of 3/2, which is uniformly distributed along the entire colloid surface, as can be seen from the first term in Eq.~(\ref{eq:FE2}). These two sources of charge conspire to a total colloidal topological charge of $+1/2$, as expected.

Finally, notice from Fig.~\ref{fig:stat}(a) and (b) that the minima of the free energy landscapes lay on lines along the poles of the colloid, \tns{which are} farthest from the transition zones. As such, the defect prefers to lie near the poles of the homeotropic (red) \bl{and} planar (blue) \tns{zones}. It follows that the colloid prefers to align its orientation either parallel or perpendicularly to the global orientation of the nematic, as depicted in Fig.~\ref{fig:solutions}. Since both minima are identical, there is no energetic preference between one state or the another. As can be seen in Fig.~\ref{fig:solutions} by comparing when the defect lies near the pole of the planar \tns{zone} (Fig.~\ref{fig:solutions}(a); blue) versus when it lies near the homeotropic pole (Fig.~\ref{fig:solutions}(c); red), bend and splay are exchanged. Since we make a one-Frank-constant approximation\tns{,} both deformations have the same energetic cost, leading to equivalent wells. In contrast, on the colloid's equator there are only local maxima, \tns{since} the defect has to overcome the repulsion of the transition zones in order to approach it. However, as can be seen in  Fig.~\ref{fig:stat}, the difference between the minima and the maxima becomes shallower as the transition zones become wider, consistent with a more \tns{distributed} negative charge. We therefore expect it to become easier for the defect to cross the equator with increased $\delta$. 

\begin{figure*}
    \includegraphics[width=\textwidth]{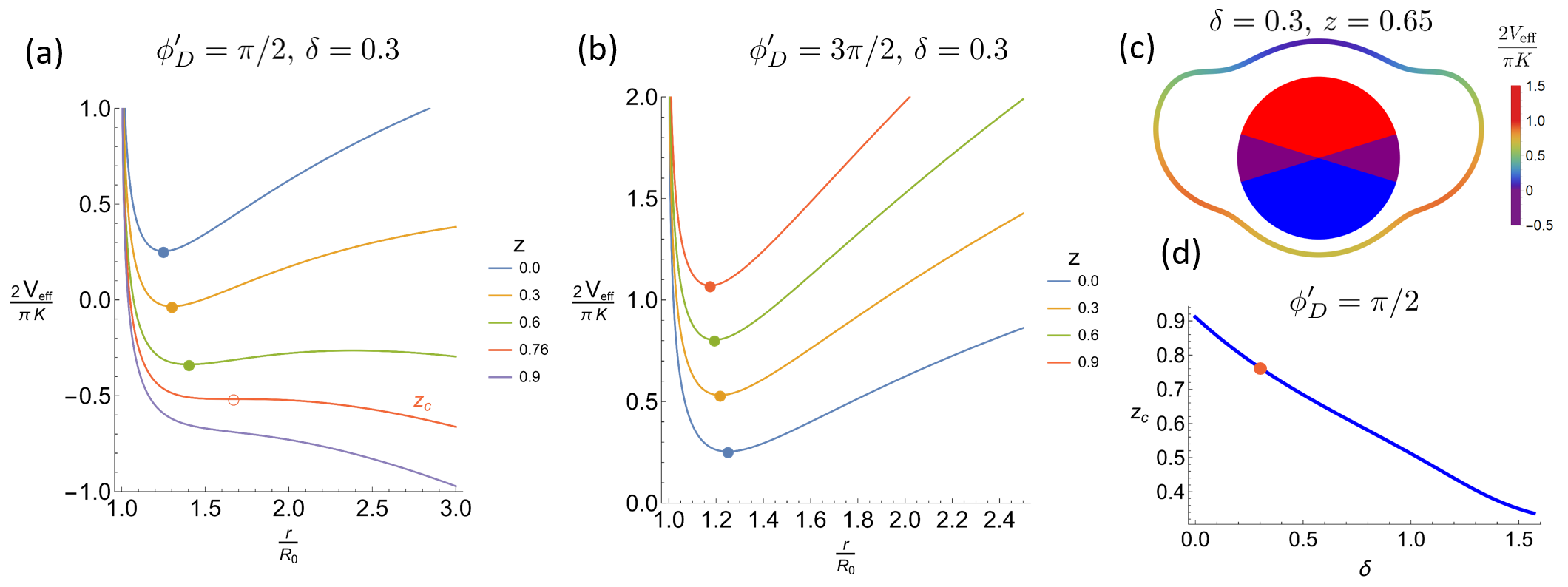}
    \captionof{figure}{(a) Effective potential for a defect sitting at the colloid's homeotropic pole (red \tns{semicircle} in Fig.~\ref{fig:solutions}) as a function of its distance to the colloid, Eq.~(\ref{eq:effective_V}), for different value of activity (parametrized through \bl{$z = 4(\alpha R_0^2\gamma’)/(K \gamma_c)$)} for $\delta=0.3$. As activity increases, the defect's proffered position (filled points) becomes larger until we reach a critical activity, $z_c$ at which there is no longer a local energy minima, (b) Same as in (a), but for a defect at the planar pole (blue \tns{semicircle} in Fig.~\ref{fig:solutions}). With increasing activity, the defect becomes closer to the colloid. There is always an energy minima. (c) Curve of stationary radii at a finite activity; defects near the homeotropic pole are closer to the colloid than those on the homeotropic pole. (d) Critical $z$ at which defects unbind from the colloid, $z_c$, as a function of the transition zone width. The orange dot marks the value of $z_c$ in (a).}
    \label{fig:stat2}
\end{figure*}

\section{Colloid propulsion}

Having established the colloid-defect interaction and the form of the nematic field in the vicinity of the Janus particle, we now seek to understand the dynamics of the colloid and its companion defect in an active \bl{nematic} fluid. 

\bl{In order to analytically approximate the propulsive force acting on the Janus colloid, we make two key assumptions. Firstly, we assume sufficiently low activity to neglect couplings between the director and the flow, which implies that the passive solutions derived in the previous section remain accurate. This approximation suggests that the active nematic length scale $\ell_a=\sqrt{K/\alpha}$ \cite{Doostmohammadi2017,Shendruk2017, Shendruk2018} is much larger than any other length scales in the system.  
Independently, we make a second assumption that the incompressible active film (of viscosity $\eta$) sits above a thin underlying oil-layer which dissipates momentum and imposes an effective friction $\gamma$. This produces a dissipative length scale $\ell_H = \sqrt{\eta/\gamma}$ \cite{Thijssen2021,martinez2021} which we take to be smaller than any other length scales in the system. This assumption amounts to taking the overdamped limit, in which frictional forces dominate over all non-active stress contributions. This has proven a useful limit for developing theoretical predictions for 2D active nematic dynamics \cite{Shankar2018,Shankar2019}. As discussed in Appendix C, the total stress is dominated by the active contribution and passive pressure contributions are negligible.

The active stress} is proportional to the nematic tensor $\boldsymbol{\sigma}_A = -\alpha\, \boldsymbol{Q}$, where $\alpha$ quantifies the activity and $\boldsymbol{Q}$ denotes the nematic tensor. \tns{Positive values of $\alpha>0$ correspond} to extensile stress, whereas $\alpha<0$ corresponds to contractile stress. 
We compute the force exerted by the active nematic on the colloid by \bl{using the total stress $\boldsymbol{\sigma}$ to compute the traction force $\boldsymbol{t}$ at the colloid's surface}
\begin{equation}
    \boldsymbol{t} = \hat{\boldsymbol{\nu}}\cdot\boldsymbol{\sigma},
\end{equation}
where $\hat{\boldsymbol{\nu}}$ denotes the normal to the colloid \bl{ surface. As seen in Appendix C, the dominant term in the total stress $\boldsymbol{\sigma}$ is its active part $\boldsymbol{\sigma}_A$. Neglecting all other passive stresses, we find}
\begin{equation}
\begin{split}
   \bl{ \boldsymbol{t}(\phi'_L) \approx}& \bl{-\alpha\cos(2(\theta_\text{BC}(\phi'_L)-\phi'_L))\,\hat{\boldsymbol{r}}}\\
    &-\alpha\sin(2(\theta_\text{BC}(\phi'_L)-\phi'_L))\,\hat{\boldsymbol{\phi}} \tns{,}
    \end{split}
\end{equation}
in which $\phi'_L$ is the polar angle with respect to the horizontal in the colloid's reference frame and $\theta_{BC}$ \bl{is given by Eq.~(\ref{eq:BC}). The total force $\boldsymbol{F}$ is obtained by integrating $\boldsymbol{t}$ over the surface of the colloid}
\begin{align}
    \label{eq:nematic_f}
    \bl{\boldsymbol{F} }&\bl{\approx} -\alpha R_0\int_0^{2 \pi}d\phi'_L\,[ \cos(2(\theta_\text{BC}-\phi'_L))\,\hat{\boldsymbol{r}} \nonumber\\
            &\qquad\qquad\qquad\qquad + \sin(2(\theta_\text{BC}-\phi'_L))\,\hat{\boldsymbol{\phi}}] \nonumber \\ 
    &= \bl{2\, \alpha\, \pi R_0\sinc\left(\delta-\frac{\pi}{2}\right)\, \hat{\boldsymbol{p}}},
\end{align}
where $\hat{\boldsymbol{p}}$ denotes the colloid's \bl{orientation, which points from the colloid center to the planar (blue) pole. Hence, for extensile activity the colloid travels towards its planar pole. In a contractile system, the colloid travels towards the homeotropic (red) pole}. There is no net-torque from the traction. While the defect position reorients the colloid (as in Fig.~\ref{fig:solutions}), the propulsion direction is always predicted to be parallel to $\hat{\boldsymbol{p}}$. 
Crucially, this shows that passive Janus particles are subject to a net force and, and thus are \tns{effectively} self-motile, when suspended in an active nematic. 
As we can see in Fig.~\ref{fig:force_mag}, the magnitude of the propulsive force increases with $\delta$ \tns{and approaches} a plateau as $\delta \rightarrow \pi/2$.  \tns{Since} this value of $\delta$ corresponds to the boundary condition of an isolated $+1/2$ defect, we see that self-propulsion of the Janus colloid with a vanishing transition zone can reach at least a 63\% of the magnitude of the ``perfect" $+1/2$ defect condition (Fig.~\ref{fig:force_mag}).

\section{Colloid-defect pair dynamics}

Similar to what occurs with a pair of two nematic topological defects, the colloid's active propulsive force can have strong effects on the dynamics of a colloid-defect pair.  In this section, we explore such phenomena by modelling both the colloid and defect as interacting point particles located at $\boldsymbol{R}_c$ and $\boldsymbol{R}_D$, respectively, \tns{in the lab frame (Fig.~\ref{fig:rotation})}. Assuming overdamped dynamics with drag coefficient $\gamma_c$ and $\gamma_D$ for the colloid and the defect, respectively, we have that their dynamics are governed by
\begin{align}
\label{eq:din_c}
    &\gamma_c \dot{\boldsymbol{R}}_c = -\boldsymbol{\nabla}_{\boldsymbol{R}_c}\mathcal{F}(\boldsymbol{R}_D-\boldsymbol{R}_c) + \boldsymbol{F},\\
    \label{eq:din_D}
     &\gamma_D \dot{\boldsymbol{R}}_D = -\boldsymbol{\nabla}_{\boldsymbol{R}_D}\mathcal{F}(\boldsymbol{R}_D-\boldsymbol{R}_c) + \boldsymbol{\xi},
\end{align}
where $\xi$ denotes random forces acting on the defect, which we assume to be negligible on the larger colloid. As such, we see that the relative coordinate $\boldsymbol{r} =\boldsymbol{R}_D-\boldsymbol{R}_c $ satisfies
\begin{equation}
     \gamma' \dot{\boldsymbol{r}} = - \boldsymbol{\nabla}_{\boldsymbol{r}}\mathcal{F}(\boldsymbol{r}) - \frac{\gamma'}{\gamma_c}\boldsymbol{F} + \frac{\gamma'}{\gamma_D}\boldsymbol{\xi},
\end{equation}
in which $\gamma' = \gamma_c\gamma_D/(\gamma_c+\gamma_D)$ is a ``reduced friction" coefficient. The radial component of the above equation has  the same structure in the colloid's \tns{co-rotational} reference frame, with the exception that\bl{, for an extensile system,}  $\boldsymbol{F}$ always points downwards \tns{in this frame (Fig.~\ref{fig:rotation})}. Thus, we write
\begin{equation}
     \gamma' \dot{r} = -\partial_r V_\text{eff}(\boldsymbol{r}) + \frac{\gamma'}{\gamma_D}\xi_r,
\end{equation}
in which we have identified the effective potential $V_\text{eff}(\boldsymbol{r}) = \mathcal{F}(\boldsymbol{r})+\frac{\gamma'}{\gamma_c}\boldsymbol{F}\cdot \boldsymbol{r}$ \tns{and the radial component of the noise $\xi_r$}. Explicitly, this amounts to
\begin{equation}
\begin{split}
\label{eq:effective_V}
      \frac{2 V_\text{eff}(\boldsymbol{r})}{K \pi} &=
    \ln(\rho)- \frac{1}{2}\ln(1-\rho^{-2})-z \sin(\phi'_D)\sinc\left(\delta-\frac{\pi}{2}\right) r\\
    &+\frac{1}{4\delta} \Im\left[\Li_2\left(\frac{e^{2 i(\phi'_D+\delta)}}{\rho^2}\right)-\Li_2\left(\frac{e^{2 i(\phi'_D-\delta)}}{\rho^2}\right)\right],
    \end{split}
\end{equation}
where we have defined the parameter $z = 4(\alpha R_0^2\gamma')/(K\gamma_c)$, which acts as a dimensionless activity number that balances the colloid size against the active nematic \tns{length scale} $\ell_a=\sqrt{K/\alpha}$ \cite{Doostmohammadi2017,Shendruk2017, Shendruk2018, Thijssen2021} and the ratio of drag coefficients $\gamma'/\gamma_c$.  As such, activity introduces a radial potential whose slope depends on the angular position of the colloid.  

This angular dependence breaks the symmetry between the homeotropic and planar poles. Configurations with defects on the poles of the homeotropic \tns{(red)} or planar \tns{(blue)} sides carry active contributions to the effective potential with opposite signs. Hence, the symmetry between the colloid being oriented parallel or perpendicular to the global nematic orientation is also broken. The reason behind this symmetry breaking is entirely due to the polarity of the self-propulsive force: \bl{In an extensile system,} when the defect sits on the homeotropic side (red), the colloidal propulsive force pulls away from the defect. This widens the separation between the two, which can be seen in Fig.~\ref{fig:stat2}(a), depicting the stationary radii shifting to larger values with increasing activity (Fig.~\ref{fig:stat2}(a); closed circles). In contrast, if the defect sits on the planar side (blue) the propulsive force on the colloid pushes it towards the defect. This reduces their separation, as can be seen from the shift of the stationary radii to smaller values with increasing activity (Fig.~\ref{fig:stat2}(b)). These effects and their variation with the angular position of the companion defect can be observed in the curve of stationary radii deformed by activity illustrated in Fig.~\ref{fig:stat2}(c).

While we have focused on extensile stresses with both $\alpha$ and $z>0$, the consequence of considering contractile forces in Fig.~\ref{fig:stat2} is straightforward. In contractile active nematics, $+1/2$ defects move in the opposite direction compared to in extensile activity and the same \tns{is to} be expected of our Janus colloid. Thus, the situation for which the colloid moves towards or away from the companion defect is flipped. The ultimate effect is simply that Fig.~\ref{fig:stat2}(a) and Fig.~\ref{fig:stat2}(b) are exchanged for a contractile activity. 

Interestingly, the effective potential for the defect on the \bl{homeotropic} pole ($\phi'_D=\pi/2$) is very similar to the one discussed in Ref.~\cite{Shankar2019}, in that the negative linear potential creates a local maximum or energy barrier which the defect can overcome through noise and thus unbind from the colloid.  However, in contrast to Ref.~\cite{Shankar2019}, our colloid has a divergent short-range repulsion, which creates an additional possibility: For a sufficiently large activity the effective potential \tns{no longer possesses} a local minimum (\emph{i.e.}, the energy barrier completely disappears), becoming effectively completely repulsive (Fig.~\ref{fig:stat2}(a); $z=\{0.76, 0.9\}$).  As a consequence, the defect will spontaneously unbind from the colloid, even in the absence of noise. The critical activity necessary for such unbinding to occur gives a $z_c$ between $0.3$ and $1$ as a function of transition zone width $\delta$ (Fig.~\ref{fig:stat2}(d)). The critical \tns{activity number} $z_c$ corresponds to an active length $\ell_a$ comparable to the colloidal diameter when the drag coefficients $\gamma_c$ and $\gamma_D$ are taken to be similar. Thus, defect unbinding is a high activity phenomena. 

At high values of activity, many of the approximations we have assumed are no longer valid such that the coupling of the nematic with the flow cannot be disregarded and the length scales at which we can assume the nematic to be aligned become comparable with the size of the colloid. Although we don't expect our analytical results to hold in such a high-activity regime, we note that the first three terms of the effective colloid-defect potential do not depend on the nematic global orientation and are sufficient to create the repulsive potential at high activity.  As such, we can expect this phenomenological prediction to persist\tns{;} if not quantitatively, at least qualitatively. To test this, future work will require full dynamical simulations of both the colloid and the active nematic.

\begin{figure}
    \includegraphics[width=\columnwidth]{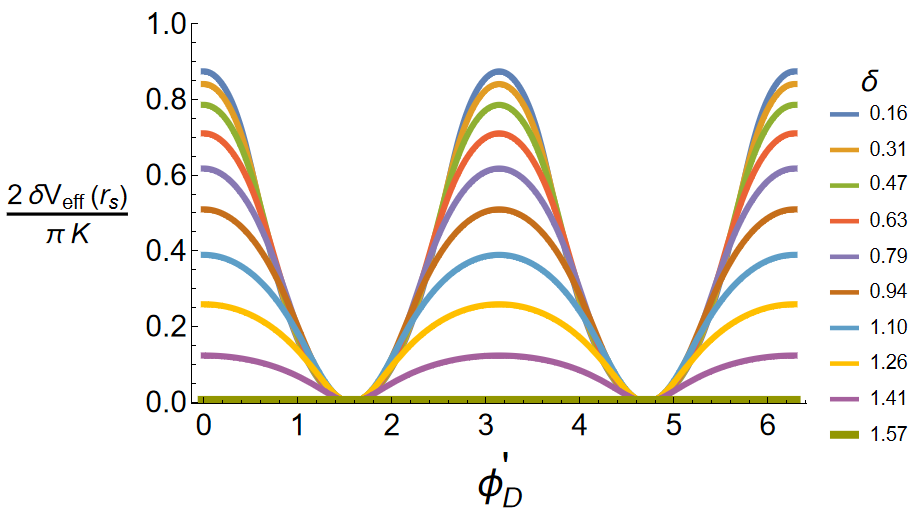}
    \caption{Effective potential ($\delta V_\text{eff}(r_s)=V_\text{eff}(r_s,\phi'_D)-V_\text{eff}(\phi'_D=\pi/2)$) along the curve of stationary radii for different \bl{angular widths of the transition zone ($\delta$)}. Notice how the energy barriers get shallower and narrow as $\delta$ increases.}
    \label{fig:barriers}
\end{figure}

In the small activity limit, in which $\ell_a \gg R_0$, or for a large discrepancy between the defect and colloidal drag coefficients, the defect should remain attached to the colloid, wiggling in the vicinity of a slightly perturbed lemon shaped ``orbit'', such as the ones in Fig.~\ref{fig:stat}. Although the probability of unbinding through noise is not zero, it is exponentially small, and thus likely to be rare.  As such, colloid and defect remain close to each other, enacting strong orientational effects on one another. In particular, we saw in the previous section that the colloid propulsion is parallel to $\hat{\boldsymbol{p}}$ in the colloid reference frame. For an extensile system, \tns{its orientation} is $\theta'_p = -\pi/2$. Relative to the global nematic alignment\tns{,} we find $\theta_p = \theta'_p+\phi_c = -\phi_D$.  As such, the direction of self-propulsion is directly dictated by the angular position of the companion defect. 

However, not all angular positions have the same probability of being occupied. On the one hand, the free energy is minimum when the position vector of the companion defects lays at the middle of either the homeotropic or planar regions. On the other hand, the maxima of the free energy occurs when the position vector of the companion defects lays along the equator of the colloid (Fig.~\ref{fig:barriers}). As a result, the companion defect lies at one of the poles of the colloid, although fluctuations may allow it to jump the energy barrier\tns{. Such jump events would be expected to sharply change} the colloid's direction of self-propulsion. The height determines the resilience of the colloid's persistence to fluctuations. As the transition zones become wider, the potential well becomes flatter and, in the case in which the entire colloid is a transition zone (\emph{i.e.}, $\delta\rightarrow \pi/2$), we observe a flat potential (Fig.~\ref{fig:barriers}). The width of the potential well on the other hand, sets how collimated the colloid trajectories are, \emph{i.e.}, how much it deviates from travelling parallel or perpendicular to the nematic director.

\section{Conclusions}

We have designed a 2D Janus boundary condition for a colloidal particle\tns{, which} becomes effectively self-propelled \tns{when immersed in an active nematic}. This boundary condition sets homeotropic anchoring \tns{on the surface of one semicircle} and planar anchoring on the opposite one, with a continuous transition zone at the equator. This boundary condition amounts to a net effective $+1/2$ topological charge.  Moreover, balancing the total stress and traction on the colloid surface reveals a non-zero net force caused by the \tns{fluid activity,} which propels the colloid forward.  Conservation of topological charge induces a companion passive $-1/2$ defect.  We \bl{analytically} solve the nematic field for an arbitrary position of this companion defect and \tns{calculate} its interaction energy with the colloid.

Moreover, we show \tns{the} angular positions of this companion defect \tns{and orientation of the colloid are linked in a one-to-one relationship}. This prediction relies on an adiabatic approximation in which the nematic\tns{'s director} and flows relax infinitely fast. However, we also show that the transition zones at the equator of the colloid have an effective topological charge density which induce an energy landscape for the position of the companion defect, with valleys at the poles and peaks at the equator. As a result, the defect tends to sit \tns{near} one of the colloid poles, which results in the colloid \tns{orientation} pointing either along the \tns{far-field nematic director} or perpendicular to it.  In the small activity regime, this leads to an anisotropic run-and-tumble-type process.

The colloid's propulsive force further affects the separation between colloid and defect, breaking the symmetry between homeotropic and planar \tns{sides} (closer at the planar side, further away on the homeotropic side).  In the high activity regime, the colloid-defect interaction becomes fully repulsive on the homeotropic side, leading to spontaneous unbinding of the companion defect. Although in such high activity regime the nematic would be highly turbulent, we expect this phenomena to persist, leading to potentially interesting dynamics.
We hope this work will serve to motivate research on predefined nematic anchoring as a method of self-propulsion in \tns{active fluids}. Similarly, we hope that it can also shed light in how to extract work from active nematics\tns{,} as well as introduce a new playground for collective phenomena.  For example, as these colloids are topologically charged they exert long-range elasticity mediated dipolar interactions which, similar to motile disclinations \cite{Shankar2019}, can lead to \tns{flocking}. Furthermore, combining these systems with light activated molecular motors \cite{Zhang2019} \tns{opens} the door to activity gradients and \tns{possible} control of \tns{colloidal} self-propulsion. 

\section{Acknowledgments}

\begin{acknowledgments}
This work was supported by the European Research Council (ERC) under the European Union’s Horizon 2020 research and innovation programme (grant agreement no. 851196). 
\end{acknowledgments}

\bibliography{Biblo}

\begin{thebibliography}{61}%
\makeatletter
\providecommand \@ifxundefined [1]{%
 \@ifx{#1\undefined}
}%
\providecommand \@ifnum [1]{%
 \ifnum #1\expandafter \@firstoftwo
 \else \expandafter \@secondoftwo
 \fi
}%
\providecommand \@ifx [1]{%
 \ifx #1\expandafter \@firstoftwo
 \else \expandafter \@secondoftwo
 \fi
}%
\providecommand \natexlab [1]{#1}%
\providecommand \enquote  [1]{``#1''}%
\providecommand \bibnamefont  [1]{#1}%
\providecommand \bibfnamefont [1]{#1}%
\providecommand \citenamefont [1]{#1}%
\providecommand \href@noop [0]{\@secondoftwo}%
\providecommand \href [0]{\begingroup \@sanitize@url \@href}%
\providecommand \@href[1]{\@@startlink{#1}\@@href}%
\providecommand \@@href[1]{\endgroup#1\@@endlink}%
\providecommand \@sanitize@url [0]{\catcode `\\12\catcode `\$12\catcode
  `\&12\catcode `\#12\catcode `\^12\catcode `\_12\catcode `\%12\relax}%
\providecommand \@@startlink[1]{}%
\providecommand \@@endlink[0]{}%
\providecommand \url  [0]{\begingroup\@sanitize@url \@url }%
\providecommand \@url [1]{\endgroup\@href {#1}{\urlprefix }}%
\providecommand \urlprefix  [0]{URL }%
\providecommand \Eprint [0]{\href }%
\providecommand \doibase [0]{https://doi.org/}%
\providecommand \selectlanguage [0]{\@gobble}%
\providecommand \bibinfo  [0]{\@secondoftwo}%
\providecommand \bibfield  [0]{\@secondoftwo}%
\providecommand \translation [1]{[#1]}%
\providecommand \BibitemOpen [0]{}%
\providecommand \bibitemStop [0]{}%
\providecommand \bibitemNoStop [0]{.\EOS\space}%
\providecommand \EOS [0]{\spacefactor3000\relax}%
\providecommand \BibitemShut  [1]{\csname bibitem#1\endcsname}%
\let\auto@bib@innerbib\@empty
\bibitem [{\citenamefont {Zhang}\ \emph {et~al.}(2021)\citenamefont {Zhang},
  \citenamefont {Mozaffari},\ and\ \citenamefont {de~Pablo}}]{Zhang2021}%
  \BibitemOpen
  \bibfield  {author} {\bibinfo {author} {\bibfnamefont {R.}~\bibnamefont
  {Zhang}}, \bibinfo {author} {\bibfnamefont {A.}~\bibnamefont {Mozaffari}},\
  and\ \bibinfo {author} {\bibfnamefont {J.~J.}\ \bibnamefont {de~Pablo}},\
  }\href {https://doi.org/10.1038/s41578-020-00272-x} {\bibfield  {journal}
  {\bibinfo  {journal} {Nature Reviews Materials}\ }\textbf {\bibinfo {volume}
  {6}},\ \bibinfo {pages} {437} (\bibinfo {year} {2021})}\BibitemShut {NoStop}%
\bibitem [{\citenamefont {Marchetti}\ \emph {et~al.}(2013)\citenamefont
  {Marchetti}, \citenamefont {Joanny}, \citenamefont {Ramaswamy}, \citenamefont
  {Liverpool}, \citenamefont {Prost}, \citenamefont {Rao},\ and\ \citenamefont
  {Simha}}]{Marchetti2013}%
  \BibitemOpen
  \bibfield  {author} {\bibinfo {author} {\bibfnamefont {M.~C.}\ \bibnamefont
  {Marchetti}}, \bibinfo {author} {\bibfnamefont {J.~F.}\ \bibnamefont
  {Joanny}}, \bibinfo {author} {\bibfnamefont {S.}~\bibnamefont {Ramaswamy}},
  \bibinfo {author} {\bibfnamefont {T.~B.}\ \bibnamefont {Liverpool}}, \bibinfo
  {author} {\bibfnamefont {J.}~\bibnamefont {Prost}}, \bibinfo {author}
  {\bibfnamefont {M.}~\bibnamefont {Rao}},\ and\ \bibinfo {author}
  {\bibfnamefont {R.~A.}\ \bibnamefont {Simha}},\ }\href
  {https://doi.org/10.1103/RevModPhys.85.1143} {\bibfield  {journal} {\bibinfo
  {journal} {Reviews of Modern Physics}\ }\textbf {\bibinfo {volume} {85}},\
  \bibinfo {pages} {1143} (\bibinfo {year} {2013})}\BibitemShut {NoStop}%
\bibitem [{\citenamefont {Dombrowski}\ \emph {et~al.}(2004)\citenamefont
  {Dombrowski}, \citenamefont {Cisneros}, \citenamefont {Chatkaew},
  \citenamefont {Goldstein},\ and\ \citenamefont {Kessler}}]{Dombrowski2004}%
  \BibitemOpen
  \bibfield  {author} {\bibinfo {author} {\bibfnamefont {C.}~\bibnamefont
  {Dombrowski}}, \bibinfo {author} {\bibfnamefont {L.}~\bibnamefont
  {Cisneros}}, \bibinfo {author} {\bibfnamefont {S.}~\bibnamefont {Chatkaew}},
  \bibinfo {author} {\bibfnamefont {R.~E.}\ \bibnamefont {Goldstein}},\ and\
  \bibinfo {author} {\bibfnamefont {J.~O.}\ \bibnamefont {Kessler}},\ }\href
  {https://doi.org/10.1103/PhysRevLett.93.098103} {\bibfield  {journal}
  {\bibinfo  {journal} {Physical Review Letters}\ }\textbf {\bibinfo {volume}
  {93}},\ \bibinfo {pages} {098103} (\bibinfo {year} {2004})}\BibitemShut
  {NoStop}%
\bibitem [{\citenamefont {Zhang}\ \emph {et~al.}(2010)\citenamefont {Zhang},
  \citenamefont {Be'er}, \citenamefont {Florin},\ and\ \citenamefont
  {Swinney}}]{Zhang2010}%
  \BibitemOpen
  \bibfield  {author} {\bibinfo {author} {\bibfnamefont {H.~P.}\ \bibnamefont
  {Zhang}}, \bibinfo {author} {\bibfnamefont {A.}~\bibnamefont {Be'er}},
  \bibinfo {author} {\bibfnamefont {E.-L.}\ \bibnamefont {Florin}},\ and\
  \bibinfo {author} {\bibfnamefont {H.~L.}\ \bibnamefont {Swinney}},\ }\href
  {https://doi.org/10.1073/pnas.1001651107} {\bibfield  {journal} {\bibinfo
  {journal} {Proceedings of the National Academy of Sciences}\ }\textbf
  {\bibinfo {volume} {107}},\ \bibinfo {pages} {13626} (\bibinfo {year}
  {2010})}\BibitemShut {NoStop}%
\bibitem [{\citenamefont {Wioland}\ \emph {et~al.}(2016)\citenamefont
  {Wioland}, \citenamefont {Lushi},\ and\ \citenamefont
  {Goldstein}}]{Wioland2016}%
  \BibitemOpen
  \bibfield  {author} {\bibinfo {author} {\bibfnamefont {H.}~\bibnamefont
  {Wioland}}, \bibinfo {author} {\bibfnamefont {E.}~\bibnamefont {Lushi}},\
  and\ \bibinfo {author} {\bibfnamefont {R.~E.}\ \bibnamefont {Goldstein}},\
  }\href {https://doi.org/10.1088/1367-2630/18/7/075002} {\bibfield  {journal}
  {\bibinfo  {journal} {New Journal of Physics}\ }\textbf {\bibinfo {volume}
  {18}},\ \bibinfo {pages} {075002} (\bibinfo {year} {2016})}\BibitemShut
  {NoStop}%
\bibitem [{\citenamefont {Serra-Picamal}\ \emph {et~al.}(2012)\citenamefont
  {Serra-Picamal}, \citenamefont {Conte}, \citenamefont {Vincent},
  \citenamefont {Anon}, \citenamefont {Tambe}, \citenamefont {Bazellieres},
  \citenamefont {Butler}, \citenamefont {Fredberg},\ and\ \citenamefont
  {Trepat}}]{Serra-Picamal2012}%
  \BibitemOpen
  \bibfield  {author} {\bibinfo {author} {\bibfnamefont {X.}~\bibnamefont
  {Serra-Picamal}}, \bibinfo {author} {\bibfnamefont {V.}~\bibnamefont
  {Conte}}, \bibinfo {author} {\bibfnamefont {R.}~\bibnamefont {Vincent}},
  \bibinfo {author} {\bibfnamefont {E.}~\bibnamefont {Anon}}, \bibinfo {author}
  {\bibfnamefont {D.~T.}\ \bibnamefont {Tambe}}, \bibinfo {author}
  {\bibfnamefont {E.}~\bibnamefont {Bazellieres}}, \bibinfo {author}
  {\bibfnamefont {J.~P.}\ \bibnamefont {Butler}}, \bibinfo {author}
  {\bibfnamefont {J.~J.}\ \bibnamefont {Fredberg}},\ and\ \bibinfo {author}
  {\bibfnamefont {X.}~\bibnamefont {Trepat}},\ }\href
  {https://doi.org/10.1038/nphys2355} {\bibfield  {journal} {\bibinfo
  {journal} {Nature Physics}\ }\textbf {\bibinfo {volume} {8}},\ \bibinfo
  {pages} {628} (\bibinfo {year} {2012})}\BibitemShut {NoStop}%
\bibitem [{\citenamefont {Blanch-Mercader}\ \emph {et~al.}(2017)\citenamefont
  {Blanch-Mercader}, \citenamefont {Vincent}, \citenamefont
  {Bazelli{\`{e}}res}, \citenamefont {Serra-Picamal}, \citenamefont {Trepat},\
  and\ \citenamefont {Casademunt}}]{Blanch-Mercader2017}%
  \BibitemOpen
  \bibfield  {author} {\bibinfo {author} {\bibfnamefont {C.}~\bibnamefont
  {Blanch-Mercader}}, \bibinfo {author} {\bibfnamefont {R.}~\bibnamefont
  {Vincent}}, \bibinfo {author} {\bibfnamefont {E.}~\bibnamefont
  {Bazelli{\`{e}}res}}, \bibinfo {author} {\bibfnamefont {X.}~\bibnamefont
  {Serra-Picamal}}, \bibinfo {author} {\bibfnamefont {X.}~\bibnamefont
  {Trepat}},\ and\ \bibinfo {author} {\bibfnamefont {J.}~\bibnamefont
  {Casademunt}},\ }\href {https://doi.org/10.1039/c6sm02188c} {\bibfield
  {journal} {\bibinfo  {journal} {Soft Matter}\ }\textbf {\bibinfo {volume}
  {13}},\ \bibinfo {pages} {1235} (\bibinfo {year} {2017})}\BibitemShut
  {NoStop}%
\bibitem [{\citenamefont {Bain}\ and\ \citenamefont
  {Bartolo}(2019)}]{Bain2019}%
  \BibitemOpen
  \bibfield  {author} {\bibinfo {author} {\bibfnamefont {N.}~\bibnamefont
  {Bain}}\ and\ \bibinfo {author} {\bibfnamefont {D.}~\bibnamefont {Bartolo}},\
  }\href {https://doi.org/10.1126/science.aat9891} {\bibfield  {journal}
  {\bibinfo  {journal} {Science}\ }\textbf {\bibinfo {volume} {363}},\ \bibinfo
  {pages} {46} (\bibinfo {year} {2019})}\BibitemShut {NoStop}%
\bibitem [{\citenamefont {Giomi}\ \emph
  {et~al.}(2013{\natexlab{a}})\citenamefont {Giomi}, \citenamefont
  {Hawley-Weld},\ and\ \citenamefont {Mahadevan}}]{Giomi2013}%
  \BibitemOpen
  \bibfield  {author} {\bibinfo {author} {\bibfnamefont {L.}~\bibnamefont
  {Giomi}}, \bibinfo {author} {\bibfnamefont {N.}~\bibnamefont {Hawley-Weld}},\
  and\ \bibinfo {author} {\bibfnamefont {L.}~\bibnamefont {Mahadevan}},\ }\href
  {https://doi.org/10.1098/rspa.2012.0637} {\bibfield  {journal} {\bibinfo
  {journal} {Proceedings of the Royal Society A: Mathematical, Physical and
  Engineering Sciences}\ }\textbf {\bibinfo {volume} {469}},\ \bibinfo {pages}
  {20120637} (\bibinfo {year} {2013}{\natexlab{a}})}\BibitemShut {NoStop}%
\bibitem [{\citenamefont {Deseigne}\ \emph {et~al.}(2010)\citenamefont
  {Deseigne}, \citenamefont {Dauchot},\ and\ \citenamefont
  {Chat{\'{e}}}}]{Deseigne2010}%
  \BibitemOpen
  \bibfield  {author} {\bibinfo {author} {\bibfnamefont {J.}~\bibnamefont
  {Deseigne}}, \bibinfo {author} {\bibfnamefont {O.}~\bibnamefont {Dauchot}},\
  and\ \bibinfo {author} {\bibfnamefont {H.}~\bibnamefont {Chat{\'{e}}}},\
  }\href {https://doi.org/10.1103/PhysRevLett.105.098001} {\bibfield  {journal}
  {\bibinfo  {journal} {Physical Review Letters}\ }\textbf {\bibinfo {volume}
  {105}},\ \bibinfo {pages} {098001} (\bibinfo {year} {2010})}\BibitemShut
  {NoStop}%
\bibitem [{\citenamefont {Paxton}\ \emph {et~al.}(2004)\citenamefont {Paxton},
  \citenamefont {Kistler}, \citenamefont {Olmeda}, \citenamefont {Sen},
  \citenamefont {{St. Angelo}}, \citenamefont {Cao}, \citenamefont {Mallouk},
  \citenamefont {Lammert},\ and\ \citenamefont {Crespi}}]{Paxton2004}%
  \BibitemOpen
  \bibfield  {author} {\bibinfo {author} {\bibfnamefont {W.~F.}\ \bibnamefont
  {Paxton}}, \bibinfo {author} {\bibfnamefont {K.~C.}\ \bibnamefont {Kistler}},
  \bibinfo {author} {\bibfnamefont {C.~C.}\ \bibnamefont {Olmeda}}, \bibinfo
  {author} {\bibfnamefont {A.}~\bibnamefont {Sen}}, \bibinfo {author}
  {\bibfnamefont {S.~K.}\ \bibnamefont {{St. Angelo}}}, \bibinfo {author}
  {\bibfnamefont {Y.}~\bibnamefont {Cao}}, \bibinfo {author} {\bibfnamefont
  {T.~E.}\ \bibnamefont {Mallouk}}, \bibinfo {author} {\bibfnamefont {P.~E.}\
  \bibnamefont {Lammert}},\ and\ \bibinfo {author} {\bibfnamefont {V.~H.}\
  \bibnamefont {Crespi}},\ }\href {https://doi.org/10.1021/ja047697z}
  {\bibfield  {journal} {\bibinfo  {journal} {Journal of the American Chemical
  Society}\ }\textbf {\bibinfo {volume} {126}},\ \bibinfo {pages} {13424}
  (\bibinfo {year} {2004})}\BibitemShut {NoStop}%
\bibitem [{\citenamefont {Liu}\ and\ \citenamefont {Sen}(2011)}]{Liu2011}%
  \BibitemOpen
  \bibfield  {author} {\bibinfo {author} {\bibfnamefont {R.}~\bibnamefont
  {Liu}}\ and\ \bibinfo {author} {\bibfnamefont {A.}~\bibnamefont {Sen}},\
  }\href {https://doi.org/10.1021/ja2082735} {\bibfield  {journal} {\bibinfo
  {journal} {Journal of the American Chemical Society}\ }\textbf {\bibinfo
  {volume} {133}},\ \bibinfo {pages} {20064} (\bibinfo {year}
  {2011})}\BibitemShut {NoStop}%
\bibitem [{\citenamefont {Valadares}\ \emph {et~al.}(2010)\citenamefont
  {Valadares}, \citenamefont {Tao}, \citenamefont {Zacharia}, \citenamefont
  {Kitaev}, \citenamefont {Galembeck}, \citenamefont {Kapral},\ and\
  \citenamefont {Ozin}}]{Valadares2010}%
  \BibitemOpen
  \bibfield  {author} {\bibinfo {author} {\bibfnamefont {L.~F.}\ \bibnamefont
  {Valadares}}, \bibinfo {author} {\bibfnamefont {Y.~G.}\ \bibnamefont {Tao}},
  \bibinfo {author} {\bibfnamefont {N.~S.}\ \bibnamefont {Zacharia}}, \bibinfo
  {author} {\bibfnamefont {V.}~\bibnamefont {Kitaev}}, \bibinfo {author}
  {\bibfnamefont {F.}~\bibnamefont {Galembeck}}, \bibinfo {author}
  {\bibfnamefont {R.}~\bibnamefont {Kapral}},\ and\ \bibinfo {author}
  {\bibfnamefont {G.~A.}\ \bibnamefont {Ozin}},\ }\href
  {https://doi.org/10.1002/smll.200901976} {\bibfield  {journal} {\bibinfo
  {journal} {Small}\ }\textbf {\bibinfo {volume} {6}},\ \bibinfo {pages} {565}
  (\bibinfo {year} {2010})}\BibitemShut {NoStop}%
\bibitem [{\citenamefont {Bricard}\ \emph {et~al.}(2013)\citenamefont
  {Bricard}, \citenamefont {Caussin}, \citenamefont {Desreumaux}, \citenamefont
  {Dauchot},\ and\ \citenamefont {Bartolo}}]{Bricard2013}%
  \BibitemOpen
  \bibfield  {author} {\bibinfo {author} {\bibfnamefont {A.}~\bibnamefont
  {Bricard}}, \bibinfo {author} {\bibfnamefont {J.~B.}\ \bibnamefont
  {Caussin}}, \bibinfo {author} {\bibfnamefont {N.}~\bibnamefont {Desreumaux}},
  \bibinfo {author} {\bibfnamefont {O.}~\bibnamefont {Dauchot}},\ and\ \bibinfo
  {author} {\bibfnamefont {D.}~\bibnamefont {Bartolo}},\ }\href
  {https://doi.org/10.1038/nature12673} {\bibfield  {journal} {\bibinfo
  {journal} {Nature}\ }\textbf {\bibinfo {volume} {503}},\ \bibinfo {pages}
  {95} (\bibinfo {year} {2013})}\BibitemShut {NoStop}%
\bibitem [{\citenamefont {Palacci}\ \emph {et~al.}(2013)\citenamefont
  {Palacci}, \citenamefont {Sacanna}, \citenamefont {Steinberg}, \citenamefont
  {Pine},\ and\ \citenamefont {Chaikin}}]{Palacci2013}%
  \BibitemOpen
  \bibfield  {author} {\bibinfo {author} {\bibfnamefont {J.}~\bibnamefont
  {Palacci}}, \bibinfo {author} {\bibfnamefont {S.}~\bibnamefont {Sacanna}},
  \bibinfo {author} {\bibfnamefont {A.~P.}\ \bibnamefont {Steinberg}}, \bibinfo
  {author} {\bibfnamefont {D.~J.}\ \bibnamefont {Pine}},\ and\ \bibinfo
  {author} {\bibfnamefont {P.~M.}\ \bibnamefont {Chaikin}},\ }\href
  {https://doi.org/10.1126/science.1230020} {\bibfield  {journal} {\bibinfo
  {journal} {Science}\ }\textbf {\bibinfo {volume} {339}},\ \bibinfo {pages}
  {936} (\bibinfo {year} {2013})}\BibitemShut {NoStop}%
\bibitem [{\citenamefont {Sokolov}\ \emph {et~al.}(2010)\citenamefont
  {Sokolov}, \citenamefont {Apodaca}, \citenamefont {Grzybowski},\ and\
  \citenamefont {Aranson}}]{Sokolov2010}%
  \BibitemOpen
  \bibfield  {author} {\bibinfo {author} {\bibfnamefont {A.}~\bibnamefont
  {Sokolov}}, \bibinfo {author} {\bibfnamefont {M.~M.}\ \bibnamefont
  {Apodaca}}, \bibinfo {author} {\bibfnamefont {B.~A.}\ \bibnamefont
  {Grzybowski}},\ and\ \bibinfo {author} {\bibfnamefont {I.~S.}\ \bibnamefont
  {Aranson}},\ }\href {https://doi.org/10.1073/pnas.0913015107} {\bibfield
  {journal} {\bibinfo  {journal} {Proceedings of the National Academy of
  Sciences of the United States of America}\ }\textbf {\bibinfo {volume}
  {107}},\ \bibinfo {pages} {969} (\bibinfo {year} {2010})}\BibitemShut
  {NoStop}%
\bibitem [{\citenamefont {Ekeh}\ \emph {et~al.}(2020)\citenamefont {Ekeh},
  \citenamefont {Cates},\ and\ \citenamefont {Fodor}}]{Ekeh2020}%
  \BibitemOpen
  \bibfield  {author} {\bibinfo {author} {\bibfnamefont {T.}~\bibnamefont
  {Ekeh}}, \bibinfo {author} {\bibfnamefont {M.~E.}\ \bibnamefont {Cates}},\
  and\ \bibinfo {author} {\bibfnamefont {{\'{E}}.}~\bibnamefont {Fodor}},\
  }\href {https://doi.org/10.1103/PhysRevE.102.010101} {\bibfield  {journal}
  {\bibinfo  {journal} {Physical Review E}\ }\textbf {\bibinfo {volume}
  {102}},\ \bibinfo {pages} {010101} (\bibinfo {year} {2020})}\BibitemShut
  {NoStop}%
\bibitem [{\citenamefont {Holubec}\ \emph {et~al.}(2020)\citenamefont
  {Holubec}, \citenamefont {Steffenoni}, \citenamefont {Falasco},\ and\
  \citenamefont {Kroy}}]{Holubec2020}%
  \BibitemOpen
  \bibfield  {author} {\bibinfo {author} {\bibfnamefont {V.}~\bibnamefont
  {Holubec}}, \bibinfo {author} {\bibfnamefont {S.}~\bibnamefont {Steffenoni}},
  \bibinfo {author} {\bibfnamefont {G.}~\bibnamefont {Falasco}},\ and\ \bibinfo
  {author} {\bibfnamefont {K.}~\bibnamefont {Kroy}},\ }\href@noop {} {\bibfield
   {journal} {\bibinfo  {journal} {Physical Review Research}\ }\textbf
  {\bibinfo {volume} {2}} (\bibinfo {year} {2020})}\BibitemShut {NoStop}%
\bibitem [{\citenamefont {Angelani}\ and\ \citenamefont {{Di
  Leonardo}}(2010)}]{Angelani2010}%
  \BibitemOpen
  \bibfield  {author} {\bibinfo {author} {\bibfnamefont {L.}~\bibnamefont
  {Angelani}}\ and\ \bibinfo {author} {\bibfnamefont {R.}~\bibnamefont {{Di
  Leonardo}}},\ }\href {https://doi.org/10.1088/1367-2630/12/11/113017}
  {\bibfield  {journal} {\bibinfo  {journal} {New Journal of Physics}\ }\textbf
  {\bibinfo {volume} {12}},\ \bibinfo {pages} {113017} (\bibinfo {year}
  {2010})}\BibitemShut {NoStop}%
\bibitem [{\citenamefont {Kaiser}\ \emph {et~al.}(2014)\citenamefont {Kaiser},
  \citenamefont {Peshkov}, \citenamefont {Sokolov}, \citenamefont {{Ten
  Hagen}}, \citenamefont {L{\"{o}}wen},\ and\ \citenamefont
  {Aranson}}]{Kaiser2014}%
  \BibitemOpen
  \bibfield  {author} {\bibinfo {author} {\bibfnamefont {A.}~\bibnamefont
  {Kaiser}}, \bibinfo {author} {\bibfnamefont {A.}~\bibnamefont {Peshkov}},
  \bibinfo {author} {\bibfnamefont {A.}~\bibnamefont {Sokolov}}, \bibinfo
  {author} {\bibfnamefont {B.}~\bibnamefont {{Ten Hagen}}}, \bibinfo {author}
  {\bibfnamefont {H.}~\bibnamefont {L{\"{o}}wen}},\ and\ \bibinfo {author}
  {\bibfnamefont {I.~S.}\ \bibnamefont {Aranson}},\ }\href
  {https://doi.org/10.1103/PhysRevLett.112.158101} {\bibfield  {journal}
  {\bibinfo  {journal} {Physical Review Letters}\ }\textbf {\bibinfo {volume}
  {112}},\ \bibinfo {pages} {158101} (\bibinfo {year} {2014})}\BibitemShut
  {NoStop}%
\bibitem [{\citenamefont {Yan}\ and\ \citenamefont {Brady}(2018)}]{Yan2018}%
  \BibitemOpen
  \bibfield  {author} {\bibinfo {author} {\bibfnamefont {W.}~\bibnamefont
  {Yan}}\ and\ \bibinfo {author} {\bibfnamefont {J.~F.}\ \bibnamefont
  {Brady}},\ }\href {https://doi.org/10.1039/c7sm01643c} {\bibfield  {journal}
  {\bibinfo  {journal} {Soft Matter}\ }\textbf {\bibinfo {volume} {14}},\
  \bibinfo {pages} {279} (\bibinfo {year} {2018})}\BibitemShut {NoStop}%
\bibitem [{\citenamefont {Pietzonka}\ \emph {et~al.}(2019)\citenamefont
  {Pietzonka}, \citenamefont {Fodor}, \citenamefont {Lohrmann}, \citenamefont
  {Cates},\ and\ \citenamefont {Seifert}}]{Pietzonka2019}%
  \BibitemOpen
  \bibfield  {author} {\bibinfo {author} {\bibfnamefont {P.}~\bibnamefont
  {Pietzonka}}, \bibinfo {author} {\bibfnamefont {{\'{E}}.}~\bibnamefont
  {Fodor}}, \bibinfo {author} {\bibfnamefont {C.}~\bibnamefont {Lohrmann}},
  \bibinfo {author} {\bibfnamefont {M.~E.}\ \bibnamefont {Cates}},\ and\
  \bibinfo {author} {\bibfnamefont {U.}~\bibnamefont {Seifert}},\ }\href
  {https://doi.org/10.1103/PhysRevX.9.041032} {\bibfield  {journal} {\bibinfo
  {journal} {Physical Review X}\ }\textbf {\bibinfo {volume} {9}},\ \bibinfo
  {pages} {041032} (\bibinfo {year} {2019})}\BibitemShut {NoStop}%
\bibitem [{\citenamefont {Ramos}\ \emph {et~al.}(2020)\citenamefont {Ramos},
  \citenamefont {Cordero},\ and\ \citenamefont {Soto}}]{Ramos2020}%
  \BibitemOpen
  \bibfield  {author} {\bibinfo {author} {\bibfnamefont {G.}~\bibnamefont
  {Ramos}}, \bibinfo {author} {\bibfnamefont {M.~L.}\ \bibnamefont {Cordero}},\
  and\ \bibinfo {author} {\bibfnamefont {R.}~\bibnamefont {Soto}},\ }\href
  {https://doi.org/10.1039/c9sm01839e} {\bibfield  {journal} {\bibinfo
  {journal} {Soft Matter}\ }\textbf {\bibinfo {volume} {16}},\ \bibinfo {pages}
  {1359} (\bibinfo {year} {2020})}\BibitemShut {NoStop}%
\bibitem [{\citenamefont {Belan}\ and\ \citenamefont
  {Kardar}(2021)}]{Belan2021}%
  \BibitemOpen
  \bibfield  {author} {\bibinfo {author} {\bibfnamefont {S.}~\bibnamefont
  {Belan}}\ and\ \bibinfo {author} {\bibfnamefont {M.}~\bibnamefont {Kardar}},\
  }\href {https://doi.org/10.1063/5.0030623} {\bibfield  {journal} {\bibinfo
  {journal} {Journal of Chemical Physics}\ }\textbf {\bibinfo {volume} {154}},\
  \bibinfo {pages} {024109} (\bibinfo {year} {2021})}\BibitemShut {NoStop}%
\bibitem [{\citenamefont {Decayeux}\ \emph {et~al.}(2021)\citenamefont
  {Decayeux}, \citenamefont {Dahirel}, \citenamefont {Jardat},\ and\
  \citenamefont {Illien}}]{Decayeux2021}%
  \BibitemOpen
  \bibfield  {author} {\bibinfo {author} {\bibfnamefont {J.}~\bibnamefont
  {Decayeux}}, \bibinfo {author} {\bibfnamefont {V.}~\bibnamefont {Dahirel}},
  \bibinfo {author} {\bibfnamefont {M.}~\bibnamefont {Jardat}},\ and\ \bibinfo
  {author} {\bibfnamefont {P.}~\bibnamefont {Illien}},\ }\href
  {http://arxiv.org/abs/2103.13244} {\bibfield  {journal} {\bibinfo  {journal}
  {arXiv:2103.13244}\ } (\bibinfo {year} {2021})}\BibitemShut {NoStop}%
\bibitem [{\citenamefont {Wu}\ and\ \citenamefont {Libchaber}(2000)}]{wu2000}%
  \BibitemOpen
  \bibfield  {author} {\bibinfo {author} {\bibfnamefont {X.-L.}\ \bibnamefont
  {Wu}}\ and\ \bibinfo {author} {\bibfnamefont {A.}~\bibnamefont {Libchaber}},\
  }\href {https://doi.org/10.1103/PhysRevLett.84.3017} {\bibfield  {journal}
  {\bibinfo  {journal} {Phys. Rev. Lett.}\ }\textbf {\bibinfo {volume} {84}},\
  \bibinfo {pages} {3017} (\bibinfo {year} {2000})}\BibitemShut {NoStop}%
\bibitem [{\citenamefont {Peng}\ \emph {et~al.}(2016)\citenamefont {Peng},
  \citenamefont {Lai}, \citenamefont {Tai}, \citenamefont {Zhang},
  \citenamefont {Xu},\ and\ \citenamefont {Cheng}}]{Peng2016}%
  \BibitemOpen
  \bibfield  {author} {\bibinfo {author} {\bibfnamefont {Y.}~\bibnamefont
  {Peng}}, \bibinfo {author} {\bibfnamefont {L.}~\bibnamefont {Lai}}, \bibinfo
  {author} {\bibfnamefont {Y.-S.}\ \bibnamefont {Tai}}, \bibinfo {author}
  {\bibfnamefont {K.}~\bibnamefont {Zhang}}, \bibinfo {author} {\bibfnamefont
  {X.}~\bibnamefont {Xu}},\ and\ \bibinfo {author} {\bibfnamefont
  {X.}~\bibnamefont {Cheng}},\ }\href
  {https://doi.org/10.1103/PhysRevLett.116.068303} {\bibfield  {journal}
  {\bibinfo  {journal} {Phys. Rev. Lett.}\ }\textbf {\bibinfo {volume} {116}},\
  \bibinfo {pages} {068303} (\bibinfo {year} {2016})}\BibitemShut {NoStop}%
\bibitem [{\citenamefont {Patteson}\ \emph {et~al.}(2016)\citenamefont
  {Patteson}, \citenamefont {Gopinath}, \citenamefont {Purohit},\ and\
  \citenamefont {Arratia}}]{Patteson2016}%
  \BibitemOpen
  \bibfield  {author} {\bibinfo {author} {\bibfnamefont {A.~E.}\ \bibnamefont
  {Patteson}}, \bibinfo {author} {\bibfnamefont {A.}~\bibnamefont {Gopinath}},
  \bibinfo {author} {\bibfnamefont {P.~K.}\ \bibnamefont {Purohit}},\ and\
  \bibinfo {author} {\bibfnamefont {P.~E.}\ \bibnamefont {Arratia}},\ }\href
  {https://doi.org/10.1039/C5SM02800K} {\bibfield  {journal} {\bibinfo
  {journal} {Soft Matter}\ }\textbf {\bibinfo {volume} {12}},\ \bibinfo {pages}
  {2365} (\bibinfo {year} {2016})}\BibitemShut {NoStop}%
\bibitem [{\citenamefont {Ortlieb}\ \emph {et~al.}(2019)\citenamefont
  {Ortlieb}, \citenamefont {Rafa\"{\i}}, \citenamefont {Peyla}, \citenamefont
  {Wagner},\ and\ \citenamefont {John}}]{Ortlieb2019}%
  \BibitemOpen
  \bibfield  {author} {\bibinfo {author} {\bibfnamefont {L.}~\bibnamefont
  {Ortlieb}}, \bibinfo {author} {\bibfnamefont {S.}~\bibnamefont {Rafa\"{\i}}},
  \bibinfo {author} {\bibfnamefont {P.}~\bibnamefont {Peyla}}, \bibinfo
  {author} {\bibfnamefont {C.}~\bibnamefont {Wagner}},\ and\ \bibinfo {author}
  {\bibfnamefont {T.}~\bibnamefont {John}},\ }\href
  {https://doi.org/10.1103/PhysRevLett.122.148101} {\bibfield  {journal}
  {\bibinfo  {journal} {Phys. Rev. Lett.}\ }\textbf {\bibinfo {volume} {122}},\
  \bibinfo {pages} {148101} (\bibinfo {year} {2019})}\BibitemShut {NoStop}%
\bibitem [{\citenamefont {Sanchez}\ \emph {et~al.}(2012)\citenamefont
  {Sanchez}, \citenamefont {Chen}, \citenamefont {DeCamp}, \citenamefont
  {Heymann},\ and\ \citenamefont {Dogic}}]{Sanchez2012}%
  \BibitemOpen
  \bibfield  {author} {\bibinfo {author} {\bibfnamefont {T.}~\bibnamefont
  {Sanchez}}, \bibinfo {author} {\bibfnamefont {D.~T.~N.}\ \bibnamefont
  {Chen}}, \bibinfo {author} {\bibfnamefont {S.~J.}\ \bibnamefont {DeCamp}},
  \bibinfo {author} {\bibfnamefont {M.}~\bibnamefont {Heymann}},\ and\ \bibinfo
  {author} {\bibfnamefont {Z.}~\bibnamefont {Dogic}},\ }\href
  {https://doi.org/10.1038/nature11591} {\bibfield  {journal} {\bibinfo
  {journal} {Nature}\ }\textbf {\bibinfo {volume} {491}},\ \bibinfo {pages}
  {431} (\bibinfo {year} {2012})}\BibitemShut {NoStop}%
\bibitem [{\citenamefont {Mallory}\ \emph {et~al.}(2014)\citenamefont
  {Mallory}, \citenamefont {Valeriani},\ and\ \citenamefont
  {Cacciuto}}]{Mallory2014}%
  \BibitemOpen
  \bibfield  {author} {\bibinfo {author} {\bibfnamefont {S.~A.}\ \bibnamefont
  {Mallory}}, \bibinfo {author} {\bibfnamefont {C.}~\bibnamefont {Valeriani}},\
  and\ \bibinfo {author} {\bibfnamefont {A.}~\bibnamefont {Cacciuto}},\ }\href
  {https://doi.org/10.1103/PhysRevE.90.032309} {\bibfield  {journal} {\bibinfo
  {journal} {Phys. Rev. E}\ }\textbf {\bibinfo {volume} {90}},\ \bibinfo
  {pages} {032309} (\bibinfo {year} {2014})}\BibitemShut {NoStop}%
\bibitem [{\citenamefont {Laskar}\ and\ \citenamefont
  {Adhikari}(2015)}]{Laskar2015}%
  \BibitemOpen
  \bibfield  {author} {\bibinfo {author} {\bibfnamefont {A.}~\bibnamefont
  {Laskar}}\ and\ \bibinfo {author} {\bibfnamefont {R.}~\bibnamefont
  {Adhikari}},\ }\href {https://doi.org/10.1039/C5SM02021B} {\bibfield
  {journal} {\bibinfo  {journal} {Soft Matter}\ }\textbf {\bibinfo {volume}
  {11}},\ \bibinfo {pages} {9073} (\bibinfo {year} {2015})}\BibitemShut
  {NoStop}%
\bibitem [{\citenamefont {Thampi}\ \emph {et~al.}(2016)\citenamefont {Thampi},
  \citenamefont {Doostmohammadi}, \citenamefont {Shendruk}, \citenamefont
  {Golestanian},\ and\ \citenamefont {Yeomans}}]{Thampi2016}%
  \BibitemOpen
  \bibfield  {author} {\bibinfo {author} {\bibfnamefont {S.~P.}\ \bibnamefont
  {Thampi}}, \bibinfo {author} {\bibfnamefont {A.}~\bibnamefont
  {Doostmohammadi}}, \bibinfo {author} {\bibfnamefont {T.~N.}\ \bibnamefont
  {Shendruk}}, \bibinfo {author} {\bibfnamefont {R.}~\bibnamefont
  {Golestanian}},\ and\ \bibinfo {author} {\bibfnamefont {J.~M.}\ \bibnamefont
  {Yeomans}},\ }\href {https://doi.org/10.1126/sciadv.1501854} {\bibfield
  {journal} {\bibinfo  {journal} {Science Advances}\ }\textbf {\bibinfo
  {volume} {2}},\ \bibinfo {pages} {1501854} (\bibinfo {year}
  {2016})}\BibitemShut {NoStop}%
\bibitem [{\citenamefont {Gruler}\ \emph {et~al.}(1999)\citenamefont {Gruler},
  \citenamefont {Dewald},\ and\ \citenamefont {Eberhardt}}]{Gruler1999}%
  \BibitemOpen
  \bibfield  {author} {\bibinfo {author} {\bibfnamefont {H.}~\bibnamefont
  {Gruler}}, \bibinfo {author} {\bibfnamefont {U.}~\bibnamefont {Dewald}},\
  and\ \bibinfo {author} {\bibfnamefont {M.}~\bibnamefont {Eberhardt}},\ }\href
  {https://doi.org/10.1007/BF03219164} {\bibfield  {journal} {\bibinfo
  {journal} {European Physical Journal B}\ }\textbf {\bibinfo {volume} {11}},\
  \bibinfo {pages} {187} (\bibinfo {year} {1999})}\BibitemShut {NoStop}%
\bibitem [{\citenamefont {Ramaswamy}\ \emph {et~al.}(2003)\citenamefont
  {Ramaswamy}, \citenamefont {Simha},\ and\ \citenamefont
  {Toner}}]{Ramaswamy2003}%
  \BibitemOpen
  \bibfield  {author} {\bibinfo {author} {\bibfnamefont {S.}~\bibnamefont
  {Ramaswamy}}, \bibinfo {author} {\bibfnamefont {R.~A.}\ \bibnamefont
  {Simha}},\ and\ \bibinfo {author} {\bibfnamefont {J.}~\bibnamefont {Toner}},\
  }\href {https://doi.org/10.1209/epl/i2003-00346-7} {\bibfield  {journal}
  {\bibinfo  {journal} {Europhysics Letters}\ }\textbf {\bibinfo {volume}
  {62}},\ \bibinfo {pages} {196} (\bibinfo {year} {2003})}\BibitemShut
  {NoStop}%
\bibitem [{\citenamefont {Ahmadi}\ \emph {et~al.}(2006)\citenamefont {Ahmadi},
  \citenamefont {Marchetti},\ and\ \citenamefont {Liverpool}}]{Ahmadi2006}%
  \BibitemOpen
  \bibfield  {author} {\bibinfo {author} {\bibfnamefont {A.}~\bibnamefont
  {Ahmadi}}, \bibinfo {author} {\bibfnamefont {M.~C.}\ \bibnamefont
  {Marchetti}},\ and\ \bibinfo {author} {\bibfnamefont {T.~B.}\ \bibnamefont
  {Liverpool}},\ }\href {https://doi.org/10.1103/PhysRevE.74.061913} {\bibfield
   {journal} {\bibinfo  {journal} {Physical Review E - Statistical, Nonlinear,
  and Soft Matter Physics}\ }\textbf {\bibinfo {volume} {74}},\ \bibinfo
  {pages} {061913} (\bibinfo {year} {2006})}\BibitemShut {NoStop}%
\bibitem [{\citenamefont {Narayan}\ \emph {et~al.}(2007)\citenamefont
  {Narayan}, \citenamefont {Ramaswamy},\ and\ \citenamefont
  {Menon}}]{Narayan2007}%
  \BibitemOpen
  \bibfield  {author} {\bibinfo {author} {\bibfnamefont {V.}~\bibnamefont
  {Narayan}}, \bibinfo {author} {\bibfnamefont {S.}~\bibnamefont {Ramaswamy}},\
  and\ \bibinfo {author} {\bibfnamefont {N.}~\bibnamefont {Menon}},\ }\href
  {https://doi.org/10.1126/science.1140414} {\bibfield  {journal} {\bibinfo
  {journal} {Science}\ }\textbf {\bibinfo {volume} {317}},\ \bibinfo {pages}
  {105} (\bibinfo {year} {2007})}\BibitemShut {NoStop}%
\bibitem [{\citenamefont {Keber}\ \emph {et~al.}(2014)\citenamefont {Keber},
  \citenamefont {Loiseau}, \citenamefont {Sanchez}, \citenamefont {DeCamp},
  \citenamefont {Giomi}, \citenamefont {Bowick}, \citenamefont {Marchetti},
  \citenamefont {Dogic},\ and\ \citenamefont {Bausch}}]{Keber2014}%
  \BibitemOpen
  \bibfield  {author} {\bibinfo {author} {\bibfnamefont {F.~C.}\ \bibnamefont
  {Keber}}, \bibinfo {author} {\bibfnamefont {E.}~\bibnamefont {Loiseau}},
  \bibinfo {author} {\bibfnamefont {T.}~\bibnamefont {Sanchez}}, \bibinfo
  {author} {\bibfnamefont {S.~J.}\ \bibnamefont {DeCamp}}, \bibinfo {author}
  {\bibfnamefont {L.}~\bibnamefont {Giomi}}, \bibinfo {author} {\bibfnamefont
  {M.~J.}\ \bibnamefont {Bowick}}, \bibinfo {author} {\bibfnamefont {M.~C.}\
  \bibnamefont {Marchetti}}, \bibinfo {author} {\bibfnamefont {Z.}~\bibnamefont
  {Dogic}},\ and\ \bibinfo {author} {\bibfnamefont {A.~R.}\ \bibnamefont
  {Bausch}},\ }\href {https://doi.org/10.1126/science.1254784} {\bibfield
  {journal} {\bibinfo  {journal} {Science}\ }\textbf {\bibinfo {volume}
  {345}},\ \bibinfo {pages} {1135} (\bibinfo {year} {2014})}\BibitemShut
  {NoStop}%
\bibitem [{\citenamefont {Kawaguchi}\ \emph {et~al.}(2017)\citenamefont
  {Kawaguchi}, \citenamefont {Kageyama},\ and\ \citenamefont
  {Sano}}]{Kawaguchi2017}%
  \BibitemOpen
  \bibfield  {author} {\bibinfo {author} {\bibfnamefont {K.}~\bibnamefont
  {Kawaguchi}}, \bibinfo {author} {\bibfnamefont {R.}~\bibnamefont
  {Kageyama}},\ and\ \bibinfo {author} {\bibfnamefont {M.}~\bibnamefont
  {Sano}},\ }\href {https://doi.org/10.1038/nature22321} {\bibfield  {journal}
  {\bibinfo  {journal} {Nature}\ }\textbf {\bibinfo {volume} {545}},\ \bibinfo
  {pages} {327} (\bibinfo {year} {2017})}\BibitemShut {NoStop}%
\bibitem [{\citenamefont {Saw}\ \emph {et~al.}(2017)\citenamefont {Saw},
  \citenamefont {Doostmohammadi}, \citenamefont {Nier}, \citenamefont
  {Kocgozlu}, \citenamefont {Thampi}, \citenamefont {Toyama}, \citenamefont
  {Marcq}, \citenamefont {Lim}, \citenamefont {Yeomans},\ and\ \citenamefont
  {Ladoux}}]{Saw2017}%
  \BibitemOpen
  \bibfield  {author} {\bibinfo {author} {\bibfnamefont {T.~B.}\ \bibnamefont
  {Saw}}, \bibinfo {author} {\bibfnamefont {A.}~\bibnamefont {Doostmohammadi}},
  \bibinfo {author} {\bibfnamefont {V.}~\bibnamefont {Nier}}, \bibinfo {author}
  {\bibfnamefont {L.}~\bibnamefont {Kocgozlu}}, \bibinfo {author}
  {\bibfnamefont {S.}~\bibnamefont {Thampi}}, \bibinfo {author} {\bibfnamefont
  {Y.}~\bibnamefont {Toyama}}, \bibinfo {author} {\bibfnamefont
  {P.}~\bibnamefont {Marcq}}, \bibinfo {author} {\bibfnamefont {C.~T.}\
  \bibnamefont {Lim}}, \bibinfo {author} {\bibfnamefont {J.~M.}\ \bibnamefont
  {Yeomans}},\ and\ \bibinfo {author} {\bibfnamefont {B.}~\bibnamefont
  {Ladoux}},\ }\href {https://doi.org/10.1038/nature21718} {\bibfield
  {journal} {\bibinfo  {journal} {Nature}\ }\textbf {\bibinfo {volume} {544}},\
  \bibinfo {pages} {212} (\bibinfo {year} {2017})}\BibitemShut {NoStop}%
\bibitem [{\citenamefont {Ellis}\ \emph {et~al.}(2018)\citenamefont {Ellis},
  \citenamefont {Pearce}, \citenamefont {Chang}, \citenamefont {Goldsztein},
  \citenamefont {Giomi},\ and\ \citenamefont {Fernandez-Nieves}}]{Ellis2018}%
  \BibitemOpen
  \bibfield  {author} {\bibinfo {author} {\bibfnamefont {P.~W.}\ \bibnamefont
  {Ellis}}, \bibinfo {author} {\bibfnamefont {D.~J.}\ \bibnamefont {Pearce}},
  \bibinfo {author} {\bibfnamefont {Y.~W.}\ \bibnamefont {Chang}}, \bibinfo
  {author} {\bibfnamefont {G.}~\bibnamefont {Goldsztein}}, \bibinfo {author}
  {\bibfnamefont {L.}~\bibnamefont {Giomi}},\ and\ \bibinfo {author}
  {\bibfnamefont {A.}~\bibnamefont {Fernandez-Nieves}},\ }\href
  {https://doi.org/10.1038/NPHYS4276} {\bibfield  {journal} {\bibinfo
  {journal} {Nature Physics}\ }\textbf {\bibinfo {volume} {14}},\ \bibinfo
  {pages} {85} (\bibinfo {year} {2018})}\BibitemShut {NoStop}%
\bibitem [{\citenamefont {Giomi}\ \emph
  {et~al.}(2013{\natexlab{b}})\citenamefont {Giomi}, \citenamefont {Bowick},
  \citenamefont {Ma},\ and\ \citenamefont {Marchetti}}]{Giomi2013a}%
  \BibitemOpen
  \bibfield  {author} {\bibinfo {author} {\bibfnamefont {L.}~\bibnamefont
  {Giomi}}, \bibinfo {author} {\bibfnamefont {M.~J.}\ \bibnamefont {Bowick}},
  \bibinfo {author} {\bibfnamefont {X.}~\bibnamefont {Ma}},\ and\ \bibinfo
  {author} {\bibfnamefont {M.~C.}\ \bibnamefont {Marchetti}},\ }\href
  {https://doi.org/10.1103/PhysRevLett.110.228101} {\bibfield  {journal}
  {\bibinfo  {journal} {Physical Review Letters}\ }\textbf {\bibinfo {volume}
  {110}},\ \bibinfo {pages} {228101} (\bibinfo {year}
  {2013}{\natexlab{b}})}\BibitemShut {NoStop}%
\bibitem [{\citenamefont {Shankar}\ \emph {et~al.}(2018)\citenamefont
  {Shankar}, \citenamefont {Ramaswamy}, \citenamefont {Marchetti},\ and\
  \citenamefont {Bowick}}]{Shankar2018}%
  \BibitemOpen
  \bibfield  {author} {\bibinfo {author} {\bibfnamefont {S.}~\bibnamefont
  {Shankar}}, \bibinfo {author} {\bibfnamefont {S.}~\bibnamefont {Ramaswamy}},
  \bibinfo {author} {\bibfnamefont {M.~C.}\ \bibnamefont {Marchetti}},\ and\
  \bibinfo {author} {\bibfnamefont {M.~J.}\ \bibnamefont {Bowick}},\ }\href
  {https://doi.org/10.1103/PhysRevLett.121.108002} {\bibfield  {journal}
  {\bibinfo  {journal} {Physical Review Letters}\ }\textbf {\bibinfo {volume}
  {121}},\ \bibinfo {pages} {108002} (\bibinfo {year} {2018})}\BibitemShut
  {NoStop}%
\bibitem [{\citenamefont {Stark}(2001)}]{Stark2001}%
  \BibitemOpen
  \bibfield  {author} {\bibinfo {author} {\bibfnamefont {H.}~\bibnamefont
  {Stark}},\ }\href {https://doi.org/10.1016/S0370-1573(00)00144-7} {\bibfield
  {journal} {\bibinfo  {journal} {Physics Reports}\ }\textbf {\bibinfo {volume}
  {351}},\ \bibinfo {pages} {387} (\bibinfo {year} {2001})}\BibitemShut
  {NoStop}%
\bibitem [{\citenamefont {Terentjev}(1995)}]{Terentjev1995}%
  \BibitemOpen
  \bibfield  {author} {\bibinfo {author} {\bibfnamefont {E.~M.}\ \bibnamefont
  {Terentjev}},\ }\href {https://doi.org/10.1103/PhysRevE.51.1330} {\bibfield
  {journal} {\bibinfo  {journal} {Physical Review E}\ }\textbf {\bibinfo
  {volume} {51}},\ \bibinfo {pages} {1330} (\bibinfo {year}
  {1995})}\BibitemShut {NoStop}%
\bibitem [{\citenamefont {Ramaswamy}\ \emph {et~al.}(1996)\citenamefont
  {Ramaswamy}, \citenamefont {Nityananda}, \citenamefont {Raghunathan},\ and\
  \citenamefont {Prost}}]{Ramaswamy1996}%
  \BibitemOpen
  \bibfield  {author} {\bibinfo {author} {\bibfnamefont {S.}~\bibnamefont
  {Ramaswamy}}, \bibinfo {author} {\bibfnamefont {R.}~\bibnamefont
  {Nityananda}}, \bibinfo {author} {\bibfnamefont {V.~A.}\ \bibnamefont
  {Raghunathan}},\ and\ \bibinfo {author} {\bibfnamefont {J.}~\bibnamefont
  {Prost}},\ }\href {https://doi.org/10.1080/10587259608034594} {\bibfield
  {journal} {\bibinfo  {journal} {Molecular Crystals and Liquid Crystals
  Science and Technology Section A: Molecular Crystals and Liquid Crystals}\
  }\textbf {\bibinfo {volume} {288}},\ \bibinfo {pages} {175} (\bibinfo {year}
  {1996})}\BibitemShut {NoStop}%
\bibitem [{\citenamefont {Dogic}\ \emph {et~al.}(2014)\citenamefont {Dogic},
  \citenamefont {Sharma},\ and\ \citenamefont {Zakhary}}]{Dogic2014}%
  \BibitemOpen
  \bibfield  {author} {\bibinfo {author} {\bibfnamefont {Z.}~\bibnamefont
  {Dogic}}, \bibinfo {author} {\bibfnamefont {P.}~\bibnamefont {Sharma}},\ and\
  \bibinfo {author} {\bibfnamefont {M.~J.}\ \bibnamefont {Zakhary}},\ }\href
  {https://doi.org/10.1146/annurev-conmatphys-031113-133827} {\bibfield
  {journal} {\bibinfo  {journal} {Annual Review of Condensed Matter Physics}\
  }\textbf {\bibinfo {volume} {5}},\ \bibinfo {pages} {137} (\bibinfo {year}
  {2014})}\BibitemShut {NoStop}%
\bibitem [{\citenamefont {Rivas}\ \emph {et~al.}(2020)\citenamefont {Rivas},
  \citenamefont {Shendruk}, \citenamefont {Henry}, \citenamefont {Reich},\ and\
  \citenamefont {Leheny}}]{Rivas2020}%
  \BibitemOpen
  \bibfield  {author} {\bibinfo {author} {\bibfnamefont {D.~P.}\ \bibnamefont
  {Rivas}}, \bibinfo {author} {\bibfnamefont {T.~N.}\ \bibnamefont {Shendruk}},
  \bibinfo {author} {\bibfnamefont {R.~R.}\ \bibnamefont {Henry}}, \bibinfo
  {author} {\bibfnamefont {D.~H.}\ \bibnamefont {Reich}},\ and\ \bibinfo
  {author} {\bibfnamefont {R.~L.}\ \bibnamefont {Leheny}},\ }\href
  {https://doi.org/10.1039/D0SM00693A} {\bibfield  {journal} {\bibinfo
  {journal} {Soft Matter}\ }\textbf {\bibinfo {volume} {16}},\ \bibinfo {pages}
  {9331} (\bibinfo {year} {2020})}\BibitemShut {NoStop}%
\bibitem [{\citenamefont {\color{black}Ebbens}\ and\ \citenamefont
  {Howse}(lack)}]{Ebbens2010}%
  \BibitemOpen
  \bibfield  {author} {\bibinfo {author} {\bibfnamefont {S.~J.}\ \bibnamefont
  {\color{black}Ebbens}}\ and\ \bibinfo {author} {\bibfnamefont {J.~R.}\
  \bibnamefont {Howse}},\ }\href {https://doi.org/10.1039/b918598d} {\bibfield
  {journal} {\bibinfo  {journal} {Soft Matter}\ }\textbf {\bibinfo {volume}
  {6}},\ \bibinfo {pages} {726} (\bibinfo {year}
  {2010\color{black}})}\BibitemShut {NoStop}%
\bibitem [{\citenamefont {Bricard}\ \emph {et~al.}(2015)\citenamefont
  {Bricard}, \citenamefont {Caussin}, \citenamefont {Das}, \citenamefont
  {Savoie}, \citenamefont {Chikkadi}, \citenamefont {Shitara}, \citenamefont
  {Chepizhko}, \citenamefont {Peruani}, \citenamefont {Saintillan},\ and\
  \citenamefont {Bartolo}}]{Bricard2015}%
  \BibitemOpen
  \bibfield  {author} {\bibinfo {author} {\bibfnamefont {A.}~\bibnamefont
  {Bricard}}, \bibinfo {author} {\bibfnamefont {J.-B.}\ \bibnamefont
  {Caussin}}, \bibinfo {author} {\bibfnamefont {D.}~\bibnamefont {Das}},
  \bibinfo {author} {\bibfnamefont {C.}~\bibnamefont {Savoie}}, \bibinfo
  {author} {\bibfnamefont {V.}~\bibnamefont {Chikkadi}}, \bibinfo {author}
  {\bibfnamefont {K.}~\bibnamefont {Shitara}}, \bibinfo {author} {\bibfnamefont
  {O.}~\bibnamefont {Chepizhko}}, \bibinfo {author} {\bibfnamefont
  {F.}~\bibnamefont {Peruani}}, \bibinfo {author} {\bibfnamefont
  {D.}~\bibnamefont {Saintillan}},\ and\ \bibinfo {author} {\bibfnamefont
  {D.}~\bibnamefont {Bartolo}},\ }\href {https://doi.org/10.1038/ncomms8470}
  {\bibfield  {journal} {\bibinfo  {journal} {Nature Communications}\ }\textbf
  {\bibinfo {volume} {6}},\ \bibinfo {pages} {7470} (\bibinfo {year}
  {2015})}\BibitemShut {NoStop}%
\bibitem [{\citenamefont {\color{black}Zhang}\ \emph
  {et~al.}(2019)\citenamefont {\color{black}Zhang}, \citenamefont {Redford},
  \citenamefont {Ruijgrok}, \citenamefont {Kumar}, \citenamefont {Mozaffari},
  \citenamefont {Zemsky}, \citenamefont {Dinner}, \citenamefont {Vitelli},
  \citenamefont {Bryant}, \citenamefont {Gardel},\ and\ \citenamefont
  {de~Pablo}}]{Zhang2019}%
  \BibitemOpen
  \bibfield  {author} {\bibinfo {author} {\bibfnamefont {R.}~\bibnamefont
  {\color{black}Zhang}}, \bibinfo {author} {\bibfnamefont {S.~A.}\ \bibnamefont
  {Redford}}, \bibinfo {author} {\bibfnamefont {P.~V.}\ \bibnamefont
  {Ruijgrok}}, \bibinfo {author} {\bibfnamefont {N.}~\bibnamefont {Kumar}},
  \bibinfo {author} {\bibfnamefont {A.}~\bibnamefont {Mozaffari}}, \bibinfo
  {author} {\bibfnamefont {S.}~\bibnamefont {Zemsky}}, \bibinfo {author}
  {\bibfnamefont {A.~R.}\ \bibnamefont {Dinner}}, \bibinfo {author}
  {\bibfnamefont {V.}~\bibnamefont {Vitelli}}, \bibinfo {author} {\bibfnamefont
  {Z.}~\bibnamefont {Bryant}}, \bibinfo {author} {\bibfnamefont {M.~L.}\
  \bibnamefont {Gardel}},\ and\ \bibinfo {author} {\bibfnamefont {J.~J.}\
  \bibnamefont {de~Pablo}},\ }\href {http://arxiv.org/abs/1912.01630}
  {\bibfield  {journal} {\bibinfo  {journal} {arXiv:1912.01630}\ } (\bibinfo
  {year} {2019})}\BibitemShut {NoStop}%
\bibitem [{\citenamefont {Vromans}\ and\ \citenamefont
  {Giomi}(2016)}]{Vromans2016}%
  \BibitemOpen
  \bibfield  {author} {\bibinfo {author} {\bibfnamefont {A.~J.}\ \bibnamefont
  {Vromans}}\ and\ \bibinfo {author} {\bibfnamefont {L.}~\bibnamefont
  {Giomi}},\ }\href {https://doi.org/10.1039/C6SM01146B} {\bibfield  {journal}
  {\bibinfo  {journal} {Soft Matter}\ }\textbf {\bibinfo {volume} {12}},\
  \bibinfo {pages} {6490} (\bibinfo {year} {2016})}\BibitemShut {NoStop}%
\bibitem [{\citenamefont {Tang}\ and\ \citenamefont
  {Selinger}(2019)}]{Tang2019}%
  \BibitemOpen
  \bibfield  {author} {\bibinfo {author} {\bibfnamefont {X.}~\bibnamefont
  {Tang}}\ and\ \bibinfo {author} {\bibfnamefont {J.~V.}\ \bibnamefont
  {Selinger}},\ }\href {https://doi.org/10.1039/C8SM01901K} {\bibfield
  {journal} {\bibinfo  {journal} {Soft Matter}\ }\textbf {\bibinfo {volume}
  {15}},\ \bibinfo {pages} {587} (\bibinfo {year} {2019})}\BibitemShut
  {NoStop}%
\bibitem [{\citenamefont {Doostmohammadi}\ \emph {et~al.}(2017)\citenamefont
  {Doostmohammadi}, \citenamefont {Shendruk}, \citenamefont {Thijssen},\ and\
  \citenamefont {Yeomans}}]{Doostmohammadi2017}%
  \BibitemOpen
  \bibfield  {author} {\bibinfo {author} {\bibfnamefont {A.}~\bibnamefont
  {Doostmohammadi}}, \bibinfo {author} {\bibfnamefont {T.~N.}\ \bibnamefont
  {Shendruk}}, \bibinfo {author} {\bibfnamefont {K.}~\bibnamefont {Thijssen}},\
  and\ \bibinfo {author} {\bibfnamefont {J.~M.}\ \bibnamefont {Yeomans}},\
  }\href {https://doi.org/10.1038/ncomms15326} {\bibfield  {journal} {\bibinfo
  {journal} {Nature Communications}\ }\textbf {\bibinfo {volume} {8}},\
  \bibinfo {pages} {1} (\bibinfo {year} {2017})}\BibitemShut {NoStop}%
\bibitem [{\citenamefont {Shendruk}\ \emph {et~al.}(2017)\citenamefont
  {Shendruk}, \citenamefont {Doostmohammadi}, \citenamefont {Thijssen},\ and\
  \citenamefont {Yeomans}}]{Shendruk2017}%
  \BibitemOpen
  \bibfield  {author} {\bibinfo {author} {\bibfnamefont {T.~N.}\ \bibnamefont
  {Shendruk}}, \bibinfo {author} {\bibfnamefont {A.}~\bibnamefont
  {Doostmohammadi}}, \bibinfo {author} {\bibfnamefont {K.}~\bibnamefont
  {Thijssen}},\ and\ \bibinfo {author} {\bibfnamefont {J.~M.}\ \bibnamefont
  {Yeomans}},\ }\href {https://doi.org/10.1039/c6sm02310j} {\bibfield
  {journal} {\bibinfo  {journal} {Soft Matter}\ }\textbf {\bibinfo {volume}
  {13}},\ \bibinfo {pages} {3853} (\bibinfo {year} {2017})}\BibitemShut
  {NoStop}%
\bibitem [{\citenamefont {Shendruk}\ \emph {et~al.}(2018)\citenamefont
  {Shendruk}, \citenamefont {Thijssen}, \citenamefont {Yeomans},\ and\
  \citenamefont {Doostmohammadi}}]{Shendruk2018}%
  \BibitemOpen
  \bibfield  {author} {\bibinfo {author} {\bibfnamefont {T.~N.}\ \bibnamefont
  {Shendruk}}, \bibinfo {author} {\bibfnamefont {K.}~\bibnamefont {Thijssen}},
  \bibinfo {author} {\bibfnamefont {J.~M.}\ \bibnamefont {Yeomans}},\ and\
  \bibinfo {author} {\bibfnamefont {A.}~\bibnamefont {Doostmohammadi}},\ }\href
  {https://doi.org/10.1103/PhysRevE.98.010601} {\bibfield  {journal} {\bibinfo
  {journal} {Physical Review E}\ }\textbf {\bibinfo {volume} {98}},\ \bibinfo
  {pages} {010601} (\bibinfo {year} {2018})}\BibitemShut {NoStop}%
\bibitem [{\citenamefont {Thijssen}\ \emph {et~al.}(2021)\citenamefont
  {Thijssen}, \citenamefont {Khaladj}, \citenamefont {Aghvami}, \citenamefont
  {Gharbi}, \citenamefont {Fraden}, \citenamefont {Yeomans}, \citenamefont
  {Hirst},\ and\ \citenamefont {Shendruk}}]{Thijssen2021}%
  \BibitemOpen
  \bibfield  {author} {\bibinfo {author} {\bibfnamefont {K.}~\bibnamefont
  {Thijssen}}, \bibinfo {author} {\bibfnamefont {D.~A.}\ \bibnamefont
  {Khaladj}}, \bibinfo {author} {\bibfnamefont {S.~A.}\ \bibnamefont
  {Aghvami}}, \bibinfo {author} {\bibfnamefont {M.~A.}\ \bibnamefont {Gharbi}},
  \bibinfo {author} {\bibfnamefont {S.}~\bibnamefont {Fraden}}, \bibinfo
  {author} {\bibfnamefont {J.~M.}\ \bibnamefont {Yeomans}}, \bibinfo {author}
  {\bibfnamefont {L.~S.}\ \bibnamefont {Hirst}},\ and\ \bibinfo {author}
  {\bibfnamefont {T.~N.}\ \bibnamefont {Shendruk}},\ }\href
  {https://www.pnas.org/content/118/38/e2106038118} {\bibfield  {journal}
  {\bibinfo  {journal} {Proceedings of the National Academy of Sciences}\
  }\textbf {\bibinfo {volume} {118}} (\bibinfo {year} {2021})}\BibitemShut
  {NoStop}%
\bibitem [{\citenamefont {Mart{\'\i}nez-Prat}\ \emph
  {et~al.}(2021)\citenamefont {Mart{\'\i}nez-Prat}, \citenamefont {Alert},
  \citenamefont {Meng}, \citenamefont {Ign{\'e}s-Mullol}, \citenamefont
  {Joanny}, \citenamefont {Casademunt}, \citenamefont {Golestanian},\ and\
  \citenamefont {Sagu{\'e}s}}]{martinez2021}%
  \BibitemOpen
  \bibfield  {author} {\bibinfo {author} {\bibfnamefont {B.}~\bibnamefont
  {Mart{\'\i}nez-Prat}}, \bibinfo {author} {\bibfnamefont {R.}~\bibnamefont
  {Alert}}, \bibinfo {author} {\bibfnamefont {F.}~\bibnamefont {Meng}},
  \bibinfo {author} {\bibfnamefont {J.}~\bibnamefont {Ign{\'e}s-Mullol}},
  \bibinfo {author} {\bibfnamefont {J.-F.}\ \bibnamefont {Joanny}}, \bibinfo
  {author} {\bibfnamefont {J.}~\bibnamefont {Casademunt}}, \bibinfo {author}
  {\bibfnamefont {R.}~\bibnamefont {Golestanian}},\ and\ \bibinfo {author}
  {\bibfnamefont {F.}~\bibnamefont {Sagu{\'e}s}},\ }\href@noop {} {\bibfield
  {journal} {\bibinfo  {journal} {arXiv preprint}\ }\textbf {\bibinfo {volume}
  {arXiv:2101.11570}} (\bibinfo {year} {2021})}\BibitemShut {NoStop}%
\bibitem [{\citenamefont {Shankar}\ and\ \citenamefont
  {Marchetti}(2019)}]{Shankar2019}%
  \BibitemOpen
  \bibfield  {author} {\bibinfo {author} {\bibfnamefont {S.}~\bibnamefont
  {Shankar}}\ and\ \bibinfo {author} {\bibfnamefont {M.~C.}\ \bibnamefont
  {Marchetti}},\ }\href {https://doi.org/10.1103/PhysRevX.9.041047} {\bibfield
  {journal} {\bibinfo  {journal} {Physical Review X}\ }\textbf {\bibinfo
  {volume} {91}},\ \bibinfo {pages} {041047} (\bibinfo {year}
  {2019})}\BibitemShut {NoStop}%
\bibitem [{\citenamefont {Logg}\ \emph {et~al.}(2012)\citenamefont {Logg},
  \citenamefont {Mardal}, \citenamefont {Wells} \emph
  {et~al.}}]{LoggMardalEtAl2012a}%
  \BibitemOpen
  \bibfield  {author} {\bibinfo {author} {\bibfnamefont {A.}~\bibnamefont
  {Logg}}, \bibinfo {author} {\bibfnamefont {K.-A.}\ \bibnamefont {Mardal}},
  \bibinfo {author} {\bibfnamefont {G.~N.}\ \bibnamefont {Wells}}, \emph
  {et~al.},\ }\href {https://doi.org/10.1007/978-3-642-23099-8} {\emph
  {\bibinfo {title} {Automated Solution of Differential Equations by the Finite
  Element Method}}}\ (\bibinfo  {publisher} {Springer},\ \bibinfo {year}
  {2012})\BibitemShut {NoStop}%
\bibitem [{\citenamefont {Aln{\ae}s}\ \emph {et~al.}(2015)\citenamefont
  {Aln{\ae}s}, \citenamefont {Blechta}, \citenamefont {Hake}, \citenamefont
  {Johansson}, \citenamefont {Kehlet}, \citenamefont {Logg}, \citenamefont
  {Richardson}, \citenamefont {Ring}, \citenamefont {Rognes},\ and\
  \citenamefont {Wells}}]{AlnaesBlechta2015a}%
  \BibitemOpen
  \bibfield  {author} {\bibinfo {author} {\bibfnamefont {M.}~\bibnamefont
  {Aln{\ae}s}}, \bibinfo {author} {\bibfnamefont {J.}~\bibnamefont {Blechta}},
  \bibinfo {author} {\bibfnamefont {J.}~\bibnamefont {Hake}}, \bibinfo {author}
  {\bibfnamefont {A.}~\bibnamefont {Johansson}}, \bibinfo {author}
  {\bibfnamefont {B.}~\bibnamefont {Kehlet}}, \bibinfo {author} {\bibfnamefont
  {A.}~\bibnamefont {Logg}}, \bibinfo {author} {\bibfnamefont {C.}~\bibnamefont
  {Richardson}}, \bibinfo {author} {\bibfnamefont {J.}~\bibnamefont {Ring}},
  \bibinfo {author} {\bibfnamefont {M.~E.}\ \bibnamefont {Rognes}},\ and\
  \bibinfo {author} {\bibfnamefont {G.~N.}\ \bibnamefont {Wells}},\ }\href
  {https://doi.org/10.11588/ans.2015.100.20553} {\bibfield  {journal} {\bibinfo
   {journal} {Archive of Numerical Software}\ }\textbf {\bibinfo {volume}
  {3}},\ \bibinfo {pages} {9} (\bibinfo {year} {2015})}\BibitemShut {NoStop}%
\end{thebibliography}%
\bibliographystyle{apsrev4-2}
\onecolumngrid
\clearpage

\twocolumngrid
\onecolumngrid

\begin{figure}[H]
    \begin{minipage}{\textwidth}
        \includegraphics[width=\textwidth]{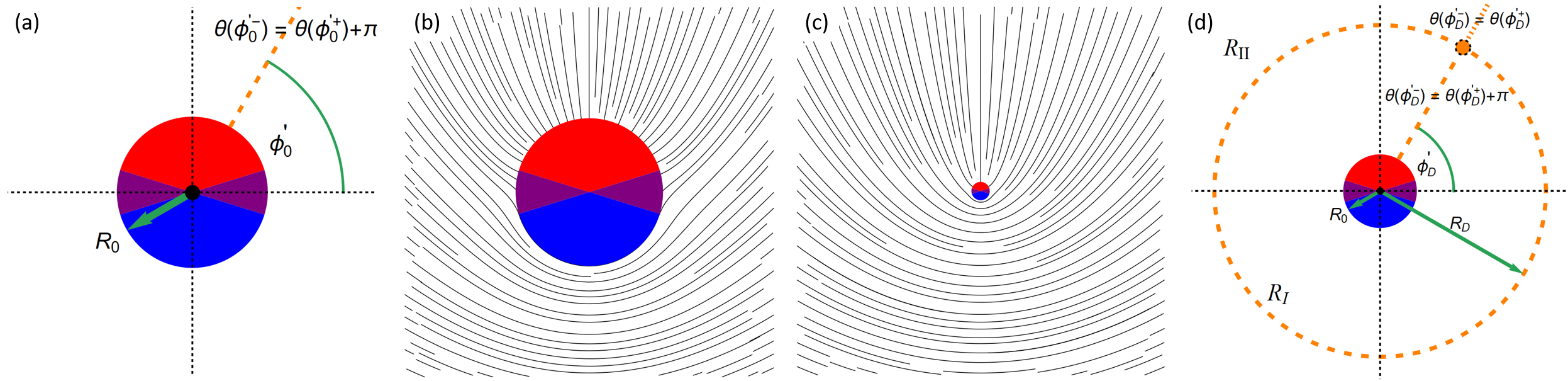}
        \captionof{figure}{(a) Non-periodic boundary conditions for an isolated colloid. Because of its topological charge, $\theta$ must jump by $\pi$ at some arbitrary radial line, (b) solution for the nematic director anchoring to the Janus colloid with its prescribed anchoring, (c) same as in (b) but from further away; at long length-scales the colloid appears exactly as a $+1/2$ defect, and (d) boundary conditions for a Janus colloid immersed in an ordered nematic. Conservation of topological charge implies the presence of a $-1/2$ defect, which lies at a radial distance $R_D$ from the colloid. For $r<R_D$, $\theta$ must jump by $\pi$ at a line whose polar angle sets the position of the defect. For $r>R_D$ the solution must be periodic. The remaining conditions are continuity and smoothness of $\theta$ at $R_D$.  }
        \label{fig:BC2}
    \end{minipage}
\end{figure}
\twocolumngrid

\appendix

\section{Appendix A: \tns{Director Field} Solution}

We focus first on the case of a Janus colloid immersed on a free nematic crystal. As described on the main text, we work in the strong anchoring condition in which the preferred colloid anchoring, Eq.~(\ref{eq:BC2}), becomes a boundary condition for the nematic.  Because this boundary condition rotates the nematic around the colloid by $\pi$, the colloid has an effective $1/2$ topological charge, which entails that after a full rotation around the colloid, $\theta$ must jump by $\pi$. Notice that this differs from the traditional boundary condition for the Laplace equation in polar coordinates which demands periodicity on the polar angle.  Here, in contrast, we must have an arbitrary line, defined by $\phi'_0$, along which we have this discontinuity in $\theta$, \emph{i.e.}, $\theta(\phi'_0-) = \theta(\phi'_0+) + \pi$, see Fig.~\ref{fig:BC2}(a). Notice that a particular solution that satisfies this boundary condition is $\theta = \phi'/2 -\pi\Theta(\phi'-\phi_0) + f(r,\phi')$, where $f(r,\phi')$ is a periodic harmonic function in $\phi$, whose role is to make sure $\theta$ satisfies the anchoring condition at the colloid's surface.  Since $\theta$ must not diverge as $r \rightarrow \infty$, then $f(r,\phi')$ must take the form $f = \sum_{n=0}^\infty(a_n \sin(n \phi')+b_n \cos(n\phi')) r^{-n}$. By demanding $\theta(R_0,\phi') =\theta_{BC}(\phi')$ we then obtain the solution
\begin{equation}
\begin{split}
    \theta(r,\phi') =& \frac{1}{2}\left(\phi'+\frac{\pi}{2}\right)-\pi\, \Theta(\phi'-\phi_0)\\
    &-\sum_{n=1}^{\infty}\frac{(1+(-1)^n) \sin(n \delta)\sin(n\phi') }{2 \delta n^2}\left(\frac{R_0}{r}\right)^n.
    \end{split}
\end{equation}
This solution is displayed on Figs.~\ref{fig:BC2}(b) and (c), which show, respectively, how the nematic follows the prescribed anchoring and how, looked from afar, the colloid behaves as a $+1/2$ topological defect.

\begin{figure}[t]
    \begin{minipage}{\textwidth}
        \includegraphics[width=\textwidth]{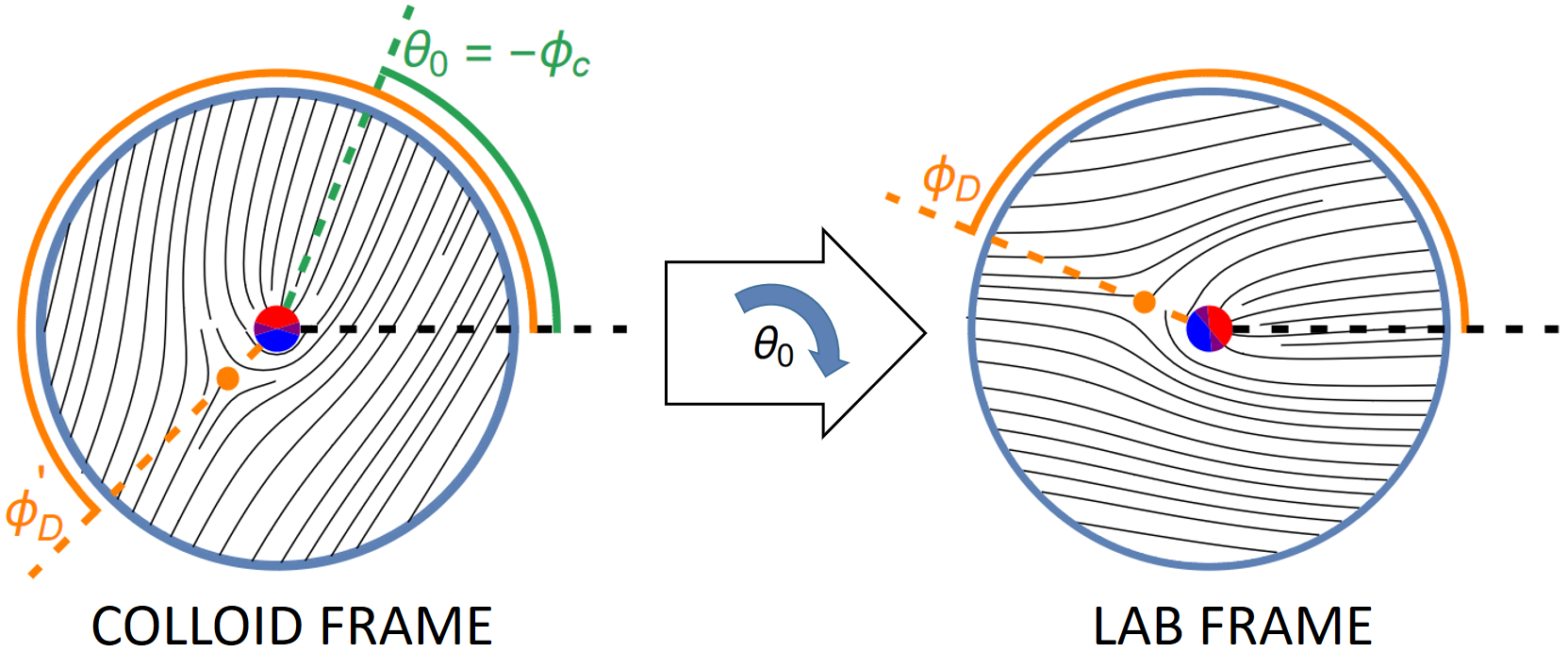}
       \bl{ \captionof{figure}{Relationship between the colloid frame and the lab frame: In the colloid frame, the colloid is always oriented with the homeotropic (red) semi-circle pointing up.  As such, the global orientation of the nematic, $\theta_0$, changes when we change the position of the defect. Since in the lab frame the nematic is oriented along the horizontal, in order to change frames we just need to rotate clock-wise by $\theta_0$.  This sets up the colloid orientation as $\phi_c=-\theta_0$.  Notice also how the angular position of the defect changes between the two frames.}
        \label{fig:rotation}}
    \end{minipage}
\end{figure}

Next, we discuss the case of a Janus colloid immersed on an ordered nematic, \emph{i.e.}, a nematic that, without loss of generality, satisfies the following asymptotic condition in the lab reference frame $\theta(r\rightarrow \infty) = 0$.  For simplicity, we work in the reference frame of the colloid in which the homeotropic \tns{semicircle (red)} points towards the $y$ axis \tns{(Fig.~\ref{fig:rotation})}.  In this reference frame, denoted by primed coordinates, the asymptotic condition becomes simply $\theta'(r\rightarrow\infty)=\theta_0$, for some constant $\theta_0$.  In order to go back to the lab frame, we simply rotate the solution by $-\theta_0$. \bl{Fig.~\ref{fig:rotation} presents a diagram of the frame transformation.} More importantly, imposing an ordered nematic far way from the colloid implies that the net topological charge in the system must be zero. However, because we know that the colloid itself has topological charge $+1/2$, this necessarily implies that there must be a companion $-1/2$ defect somewhere in the system, say at a radial distance $R_D$ and a polar angle $\phi_D$.  This naturally divides our system into two regions, see Fig.~\ref{fig:BC2}(d). On the one hand, for $R_0 \leq r < R_D$ we have a situation very similar to the colloid in the free nematic: Any rotation around the colloid must lead to a jump in $\theta'$ of $\pi$.  As such, we use the same boundary condition as before with the caveat that in this case $\phi'_0 = \phi'_D$.  That is, the line that in the previous case was arbitrary, in this case determines the angular position of the defect. Therefore, we look for a solution of the form $\theta'_I(r,\phi') = 1/2\phi'-\pi \Theta(\phi'-\phi'_D)+\sum_{n=0}^\infty(a_n \sin(n\phi')r^n+b_n \sin(n\phi')r^{-n}+c_n \cos(n\phi')r^{n}+d_n \cos(n\phi')r^{-n})$, where now, because this region is finite, we also allow for positive powers of $r$. On the other hand, for $r>R_D$ the total topological charge encircled by a loop is zero, which implies that our solution should be periodic in $\phi$; \emph{i.e.}, $\theta'(r>R_D,\phi'^{-}_D) =  \theta'(r>R_D,\phi'^{+}_D)$. As such, the solution in this region must have the form $\theta'_{II}(r,\phi') = \sum_{n=0}^\infty(A_n\sin(n\phi')+B_n\sin(n\phi'))r^{-n}$. The coefficients in both $\theta'_I$ and $\theta'_{II}$ are obtained by imposing the colloid anchoring ($\theta'_I(R_0,\phi') = \theta_{BC}(\phi')$), continuity ($\theta'_I(R_D,\phi')=\theta'_{II}(R_D,\phi')$) and smoothness ($\partial_{r}\theta'_I(r,\phi')|_{r=R_D}=\partial_{r}\theta'_{II}(r,\phi')|_{r=R_D}$). With this, we obtain the solution
\begin{widetext}
\begin{align}
    \begin{split}
        \theta'_I(r,\phi') = {}&  \frac{1}{2}\left(\phi'+\frac{\pi}{2}\right)-\pi \Theta(\phi'-\phi'_D)\\
        &-\sum_{n=1}^{\infty}\left[\frac{(1+(-1)^n) \sin(n \delta)\sin(n\phi') }{2 \delta n^2}\left(\frac{R_0}{r}\right)^n\right.
        \left.+\frac{\sin(n(\phi'-\phi'_D))}{2 n}\bl{\left\{\left(\frac{R_0^2}{R_D r}\right)^n-\left(\frac{r}{R_D}\right)^n\right\}}\right],
    \end{split}\\
    \begin{split}
        \theta'_{II}(r,\phi')  = {}& \frac{1}{4}(2\phi'_D-\pi)\\
       &\bl{-\sum_{n=1}^{\infty}\left[\frac{(1+(-1)^n) \sin(n \delta)\sin(n\phi') }{2 \delta n^2}\left(\frac{R_0}{r}\right)^n\right.
        \left.+\frac{\sin(n(\phi'-\phi'_D))}{2 n}\left\{\left(\frac{R_0^2}{R_D r}\right)^n+\left(\frac{R_D}{r}\right)^n\right\}\right]}.
    \end{split}\label{eq:theta_II}
\end{align}
\end{widetext}
\onecolumngrid

\twocolumngrid
The above solution determines that $\theta_0 = (2\phi'_D-\pi)/4$ and thus that the orientation of the colloid is given by $\phi_c = -\theta_0$. Rotating by $-\theta_0$, \emph{i.e.}, $\theta(r,\phi) = \theta'(r,\phi+\theta_0)-\theta_0$, finally leads us to the expression in the main text, Eqs.~(\ref{eq:solution1}-\ref{eq:solution2}).

\section{Appendix B: Computation of Free Energy}
\label{appendix:Free_Energy}

\begin{figure}[b]
    \includegraphics[width=\columnwidth]{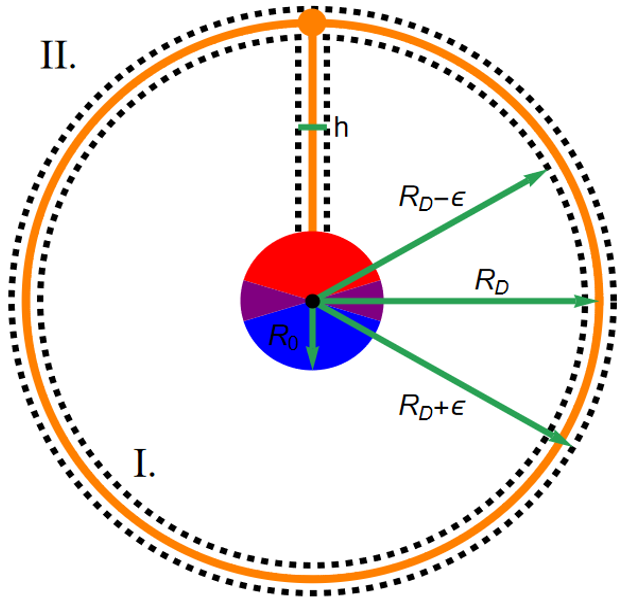}
    \caption{Diagram of the regularization scheme. We remove the defect from the region of integration by excluding an annular region of radial width $2\epsilon$, here depicted with dashed lines. Similarly, we also exclude the rectangle of width $h$, also in dashed lines,  containing the line that joins the defect with the colloid. }
    \label{fig:regularization}
\end{figure}

In this section we compute the free energy of the system for a defect at $R_D$, $\phi_D$ using Eq.~\ref{free_energy}. Throughout this entire section, all coordinates are with respect to the colloid's reference frame: For simplicity we drop the primes. Notice that this computation requires some care, as the presence of the companion defect makes the integral divergent. As such, we must first regularize and then renormalize this expression. To do the former, we constrain the region of integration by excluding an annulus of inner radius $R_D-\epsilon$ and outer radius $R_D+\epsilon$  centered in the colloid. This allows us to approach the defect in a controlled way by taking the limit $\epsilon\rightarrow 0$. In addition, we also exclude the line joining the colloid with the defect, taking out a sector of width $h$, after which we take the limit $h\rightarrow 0$. In practical terms, this amounts to ignore the squared of the derivative of the \tns{Heaviside} function inside $\theta_{I}$, which is not defined. See Fig.~\ref{fig:regularization} for a depiction of the region of integration.

By performing the integral in the regularized domain , using that $\text{Li}_s(z) = \sum_{k=1}^\infty\frac{z^k}{k^s}$, where $\text{Li}_s(z)$ is the \tns{poly-logarithm} of order $s$, and taking the limit $\epsilon/R_D \ll 1$ we obtain that
\begin{equation}
\begin{split}
    \frac{2 \mathcal{F}}{\pi K} = & -\frac{\ln(2)}{2}-\frac{1}{2}\ln\left(\frac{\epsilon}{R_0}\right)+\ln(\rho)-\frac{1}{2}\ln\left(1-\frac{1}{\rho^2}\right)\\
    &+\frac{1}{4\delta} \Im\left[\Li_2\left(\frac{e^{2 i(\phi_D+\delta)}}{\rho^2}\right)-\Li_2\left(\frac{e^{2 i(\phi_D-\delta)}}{\rho^2}\right)\right]\\
    &+\frac{1}{16\delta^2} \Re\left[\Li_3(e^{4i\delta})-\zeta(3)\right],
    \end{split}
\end{equation}
which has allowed us to isolate the divergent part of the integral in the $\ln(\epsilon/R_0)$ term.  We now proceed to renormalize by \tns{subtracting} the bare energy $2\mathcal{F}_0/(\pi K) = -\ln(2)/2-\ln(\epsilon/R_0)/2+Re[\Li_3(e^{4i\delta})-\zeta(3)]/(16 \delta^2)$, which leads to the expression on the main text, Eq.~(\ref{eq:FE1}).

Finally, in order to go from Eq.~(\ref{eq:FE1}) to Eq.~(\ref{eq:FE2}), we use the identity 
\begin{equation}
    \text{Li}_2(a) = \int_0^a dt \frac{\ln(1-t)}{t}. 
\end{equation}
We then change the variable of integration to $\beta$, defined via $t=e^{2i(\phi_D+\beta)/\rho^2}$, to find

\begin{widetext}
\begin{equation}
\begin{split}
    \Li_2\left(\frac{e^{2 i(\phi_D+\delta)}}{\rho^2}\right)-\Li_2\left(\frac{e^{2 i(\phi_D-\delta)}}{\rho^2}\right)
    &=-2 i \int_{-\delta}^{\delta} d\beta \ln\left(1-\frac{e^{2 i (\phi_D+\beta)}}{\rho^2}\right)\\
    &=-2i\left(\int_{-\delta}^{\delta}d\beta \ln\left[1-\frac{e^{i(\phi_D-\beta)}}{\rho}\right]+\int_{\pi-\delta}^{\pi+\delta}d\beta \ln\left[1-\frac{e^{i(\phi_D-\beta)}}{\rho}\right]\right),
\end{split}
\end{equation}
\end{widetext}
from which Eq.~(\ref{eq:FE2}) naturally follows.

\section{Appendix C: Neglecting hydrodynamic stresses and pressure}

In the main text, leading up to Eq.~(\ref{eq:nematic_f}), we neglected all hydrodynamic stresses, effectively considering overdamped active stresses. 
In this limit, the viscous and passive nematohydrodynamic stresses can be neglected in the limit of a vanishingly small hydrodynamic length $\ell_H$. Yet, the hydrostatic pressure may not be necessarily small; however, we demonstrate here that it is indeed negligible. 

The passive pressure $P$ contributes to the force on the colloid 
\begin{align}
    \label{eq:pressure_f}
    \begin{split}
        \boldsymbol{F}_P =  -R_0\int_0^{2 \pi}d\phi'_L\, P(\phi'_L)\hat{\boldsymbol{r}}.
    \end{split}
\end{align}
Unfortunately, the pressure contribution cannot be obtained analytically. To numerically estimate the pressure, we write the relevant Stokes equation for the fluid velocity
\begin{equation}
\label{eq:simplified_flow}
-\gamma \boldsymbol{v} - \boldsymbol{\nabla} P - \alpha \boldsymbol{\nabla}\cdot\boldsymbol{Q} =0,
\end{equation}
which is complemented by the incompressibility condition $\boldsymbol{\nabla}\cdot\boldsymbol{v}=0$. Using incompressibility in Eq.~(\ref{eq:simplified_flow}), leads to
\begin{equation}
\label{eq:poisson}
    \nabla^2 P = -\alpha \boldsymbol{\nabla}\boldsymbol{\nabla}:\boldsymbol{Q}.
\end{equation}
The boundary condition associated to this equation comes from demanding an impenetrable colloid, \emph{i.e.}, $\hat{\boldsymbol{r}}\cdot\boldsymbol{v} = 0$.  This leads us to the following Neumann boundary condition for the pressure
\begin{equation}
\label{eq:neumann}
    \hat{\boldsymbol{r}}\cdot\nabla P = -\alpha\, \hat{\boldsymbol{r}}\cdot(\boldsymbol{\nabla}\cdot \boldsymbol{Q}).
\end{equation}
We solve Eqs.~(\ref{eq:poisson}) and (\ref{eq:neumann}) using finite elements \cite{LoggMardalEtAl2012a, AlnaesBlechta2015a} and use the solution in Eq.~(\ref{eq:pressure_f}) to obtain an estimate for $\boldsymbol{F}_P$.  The results show that, although this force is different from zero, it is much smaller than the active contribution as it satisfies $|\boldsymbol{F}_P|/|\boldsymbol{F}_A| \sim 10^{-2}$, allowing us to neglect it.


\end{document}